\documentclass[useAMS,usenatbib]{mn2e}
\usepackage{graphicx}
\usepackage{multicol}
\usepackage{amssymb}
\usepackage{amsmath}
\usepackage{caption}
\usepackage{subfig}
\usepackage{courier}
\usepackage{placeins}
\title[Interstellar gas towards HESS J1640$-$465 $\&$ HESS J1641$-$463]{Interstellar gas towards the TeV gamma-ray sources \mbox{HESS J1640$-$465} and HESS J1641$-$463}

\author[J. C. Lau et al.]{J. C. Lau$^{1}$\thanks{E-mail:james.lau@adelaide.edu.au}, G. Rowell$^{1}$, M. G. Burton$^{2,3}$, Y. Fukui$^{4}$, F. A. Aharonian$^{5,6,7}$, \and I. Oya$^{8}$, J. Vink$^{9}$, S. Ohm$^{8}$, and S. Casanova$^{5,10}$\\
$^{1}$School of Physical Sciences, University of Adelaide, Adelaide, SA 5005, Australia\\
$^{2}$School of Physics, University of New South Wales, Sydney, NSW 2052, Australia\\
$^{3}$Armagh Observatory and Planetarium, College Hill, Armagh, BT61 9DG, Northern Ireland, UK\\
$^{4}$Department of Physics, University of Nagoya, Furo-cho, Chikusa-ku, Nagoya, 464-8601, Japan\\
$^{5}$Max-Planck-Institut f\"{u}r Kernphysik, PO Box 103980, D-69029 Heidelberg, Germany\\
$^{6}$Dublin Institute for Advanced Studies, 31 Fitzwilliam Place, Dublin 2, Ireland\\
$^{7}$National Research Nuclear University (MEPHI), 115409, Moscow, Russia\\
$^{8}$DESY, D-15738 Zeuthen, Germany\\
$^{9}$GRAPPA, Anton Pannekoek Institute for Astronomy, University of Amsterdam, Science Park 904, 1098
XH Amsterdam,\\The Netherlands\\
$^{10}$Instytut Fizyki Ja¸drowej PAN, ul. Radzikowskiego 152, 31-342 Krakow, Poland\\}

\begin{document}

\pagerange{\pageref{firstpage}--\pageref{lastpage}} \pubyear{2016}

\maketitle

\label{firstpage}

\begin{abstract}
We present a detailed analysis of the interstellar medium towards the TeV $\gamma$-ray sources HESS\,J1640$-$465 and HESS\,J1641$-$463 using results from the Mopra Southern Galactic Plane CO Survey and from a Mopra 7 mm-wavelength study. The $\gamma$-ray sources are positionally coincident with two supernova remnants G338.3$-$0.0 and G338.5+0.1 respectively. A bright complex of HII regions connect the two SNRs and TeV objects. Observations in the CO(1-0) transition lines reveal substantial amounts of diffuse gas positionally coincident with the $\gamma$-ray sources at multiple velocities along the line of sight, while 7 mm observations in CS, SiO, HC$_{3}$N and CH$_{3}$OH transition lines reveal regions of dense, shocked gas. Archival HI data from the Southern Galactic Plane Survey was used to account for the diffuse atomic gas. Physical parameters of the gas towards the TeV sources were calculated from the data. We find that for a hadronic origin for the $\gamma$-ray emission, the cosmic-ray enhancement rates are $\sim 10^{3}$ and $10^{2}$ times the local solar value for HESS\,J1640$-$465 and HESS\,J1641$-$463 respectively.

\end{abstract}


\section{Introduction}
HESS\,J1640$-$465 and HESS\,J1641$-$463 are two adjacent and intriguing very-high-energy (VHE, E $>$ 100 GeV) $\gamma$-ray sources whose origins are uncertain. Knowledge of the distribution of interstellar gas towards these sources is vital in order to differentiate between possible models of TeV $\gamma$-ray production. In particular, understanding the hadronic production model of TeV gamma-rays in which highly accelerated cosmic rays (CRs) interact with target atomic and molecular gas.

HESS\,J1640$-$465 is a VHE $\gamma$-ray source first discovered  by the High Energy Stereoscopic System (H.E.S.S.) during a survey of the Galactic Plane \citep{hess_plane}. It is positionally coincident with the supernova remnant G338.3$-$0.0 \citep{1996A&AS..118..329W}. Observations with XMM-Newton detected a slightly extended and asymmetric X-ray source towards the geometric centre of the SNR G338.3$-$0.0 \citep{2007ApJ...662..517F}. Follow up observations with Chandra in X-rays revealed an extended nebula with a point-like source, a possible associated pulsar \citep{2009ApJ...706.1269L}. It was suggested that the X-rays and VHE $\gamma$-rays were then due to synchrotron and inverse-Compton  emission from a pulsar wind nebula (PWN). Multi-frequency radio analysis by \cite{110.5301v1} placed upper limits on the radio flux from the region of the supposed PWN. Observations with the Fermi Large Area Telescope (Fermi-LAT) revealed a high energy (HE) $\gamma$-ray source (1FGL 1640.8-4634) coincident with HESS\,J1640$-$465 \citep{004.2936v2}.

Further observations by H.E.S.S. \citep{2014MNRAS.439.2828A} show that the VHE $\gamma$-rays seen from HESS\,J1640$-$465 overlap significantly with the SNR shell of G338.3$-$0.0. The VHE $\gamma$-ray spectrum connects smoothly with the GeV $\gamma$-ray spectrum obtained by the analysis of five years worth of Fermi-LAT data towards HESS\,J1640$-$465 \citep{2014ApJ...794L..16L}. The smooth, flat, and featureless $\gamma$-ray spectrum strengthened the hadronic scenario in which CRs accelerated by the SNR are interacting with nearby gas \citep{2014MNRAS.439.2828A,2014ApJ...794L..16L}. However, a contribution to the detected flux by a pulsar or PWN could not be ruled out. Recent work by \cite{2016A&A...589A..51S} looked at the SED of HESS\,J1640$-$465 using latest data from H.E.S.S. and Fermi-LAT observations, together with an updated hadronic $\gamma$-ray model \citep{2014PhRvD..90l3014K} and archival atomic (HI) and molecular data. Their fit yielded a spectra index $\Gamma = 2.13$ with a cutoff energy $E_{\text{cut}}=54$ TeV. This hadronic model was found to fully describe the $\gamma$-ray spectrum of HESS\,J1640$-$465.
 
Shortly after the H.E.S.S. publication \citep{2014MNRAS.439.2828A}, \cite{2014ApJ...788..155G} discovered pulsed X-ray emission using the Nuclear Spectroscopic Telescope Array (NuSTAR) emanating from the previously discovered X-ray source seen towards the centre of HESS\,J1640$-$465. The newly discovered pulsar PSR J1640$-$4631 has a period of 206 ms with a spin-down luminosity of $4.4 \times 10^{36}$ erg s$^{-1}$  and a characteristic age of 3350 years. Modelling of leptonic $\gamma$-ray production suggested that a PWN could be responsible for the TeV emission from HESS\,J1640$-$465, although fine-tuning is required to explain the smooth GeV and TeV spectrum. 

HESS\,J1641$-$463 was initially unnoticed by standard H.E.S.S. detection techniques due to its low brightness and proximity to the bright source HESS\,J1640$-$465. Energy cuts and deeper observations led to the positive identification of the new TeV $\gamma$-ray source at a significance of 8.5$\sigma$ at energies above 4 TeV \citep{2014ApJ...794L...1A}. HESS\,J1641$-$463 has an unusually hard spectrum (photon index $\Gamma \approx 2$) with no obvious sign of a cut-off. Analysis of Fermi-LAT data \citep{2014ApJ...794L..16L} reported the detection of 2 distinct GeV sources positionally coincident with HESS\,J1641$-$463 and the nearby HESS\,J1640$-$465. HESS\,J1641$-$463 is positionally coincident with the radio SNR G338.5+0.1. The SNR itself is seen as a poorly defined circle of non-thermal emission \citep{1996A&AS..118..329W}. The pair of SNRs seen towards the two H.E.S.S. sources are connected by a complex of HII regions, which includes G338.4+0.0 and G338.45+0.06.

The production of TeV $\gamma$-ray emission from HESS\,J1641$-$463 via leptonic processes via a population of electrons with energies of several hundred TeV up-scattering background photons was considered by \cite{2014ApJ...794L...1A}. These electrons could be sourced from the coincident SNR G338.5+0.1 or even from a nearby PWN. This leptonic scenario, however, should be accompanied by a characteristic break in the $\gamma$-ray spectrum at multi-TeV energies resulting from the Klein-Nishina effect on the cross-section for inverse-Compton scattering at high energies. The lack of such a characteristic break in the $\gamma$-ray spectrum of HESS\,J1641$-$463 led the authors to disfavour the leptonic scenario.

A more promising scenario is that the $\gamma$-ray emission from HESS\,J1641$-$463 is due to interstellar medium (ISM) illuminated by highly accelerated CRs. Modelling by \cite{2014ApJ...794L...1A} indicate that the TeV spectrum of HESS\,J1641$-$463 could be produced by distribution of protons (with a power-law slope of $-2.1$) interacting with molecular gas seen by using CO(1-0) data taken with the Nanten radio telescope. The proton spectrum would need to have a high cut-off energy ($>100$ TeV) and represents one of the hardest spectra associated with a TeV $\gamma$-ray source extending into the PeV energy range; a so called PeVatron. The coincident SNR G338.5+0.1 could possibly be the source of these CR protons provided it had a young age ($\lesssim1$ kyr), as the proton spectrum agrees with predictions of diffusive shock acceleration in young SNRs. However an older SNR (5-17 kyr, \citealt{2014ApJ...794L...1A}) would not be able to accelerate CRs up to PeV energies \citep{2013MNRAS.431..415B}, and would require another CR source. An intriguing idea is that VHE protons accelerated by the young SNR coincident with HESS\,J1640$-$465, SNR G338.3$-$0.0 (with an age of 1-2 kyr to 5-8 kyr \citep{004.2936v2,2014MNRAS.439.2828A}, could be diffusively reaching the gas towards HESS\,J1641$-$463. The energy-dependant process of diffusion would preferentially allow higher energy CRs to reach the target material earlier \citep{1996A&A...309..917A}, producing the hard proton spectrum that is needed to generate the TeV $\gamma$-ray spectrum of HESS\,J1641$-$463.

Another puzzling aspect about HESS\,J1641$-$463 is the marked difference between the GeV and TeV components of its $\gamma$-ray spectrum. The GeV spectrum as measured by Fermi-LAT is very soft, which is in stark contrast to the very hard TeV spectrum as measured by H.E.S.S. This suggests that there may be two different sources to the GeV and TeV components. A possible scenario would be GeV $\gamma$-rays are produced by less energetic CRs from the old SNR G338.5+0.1 illuminating ambient gas, with TeV emission produced by higher energy CRs from the younger SNR G338.3$-$0.0.

Any attempt to fully understand the origin scenarios of both HESS\,J1640$-$465 and HESS\,J1641$-$463 requires a detailed understanding of the distribution of the interstellar medium in the surrounding environment. Thus we have used high-resolution data collected by the Mopra radio telescope in this study. As part of the Mopra Southern Galactic Plane CO Survey, the distribution of diffuse ($\overline{n}$ $\lesssim 10^{3}$ cm$^{-3}$) interstellar medium was traced towards HESS\,J1640$-$465 and HESS\,J1641$-$463. In addition we have taken complimentary data in the 7 mm wavelength band, targeting the dense ($\overline{n} \gtrsim 10^{4}$ cm$^{-3}$) gas tracer CS(1-0) as well as the tracers SiO(1-0), CH$_{3}$OH and HC$_{3}$N.

In \S 2 we describe the parameters of the data taken by the Mopra radio telescope and the data reduction processes. The gas parameter calculations we apply to these data are described in \S \ref{sec:calc}. In \S \ref{sec:results} we investigate the gas distribution towards HESS\,J1640$-$465 and HESS\,J1641$-$463 and in \S \ref{sec:discussion} we discuss the impact our results have on the possible emission scenarios for the TeV sources.

\subsection{Distance to SNRs and HII complex}
\label{sec:distance}
The distance to SNR G338.3$-$0.0 and SNR G338.5+0.1, and the HII complex containing G338.4+0.0 and G338.45+0.06 have previously been reported in several studies utilising observations in the 21 cm HI spectral line.

\cite{2009ApJ...706.1269L} derived a distance of 8 - 13 kpc for SNR G338.3$-$0.0 and the HII surrounding region based on HI absorption features. This is in agreement with previous work presented by \cite{2007A&A...468..993K}, who also used HI absorption to derive a distance of $11.7^{+2.0}_{-0.5}$ kpc for G338.4+0.0. The nearby SNR G338.5+0.1, coincident with HESS\,J1641$-$463, was found to have a very similar HI absorption profile as G338.4+0.0, which led to the assertion that it too was most likely located at $\sim$11 kpc.

The velocity along the line-of-sight (v$_{\text{LSR}}$) of the HII regions in the complex have been measured to have values of $\sim-40$ to $-30$ km/s \citep{1987A&A...171..261C,2003A&A...397..133R,2012MNRAS.420.1656U}. The kinematic distance ambiguities towards these HII regions have been addressed, and they have been constrained to the far distance \citep{2012MNRAS.420.1656U}.

This places the two SNRs and the HII complex in the Norma II spiral arm at the far side of the Galaxy. In our results in \S4, we have used the Galactic rotation curve in \cite{2007A&A...468..993K} to calculate distances for corresponding v$_{\text{LSR}}$. For purposes of our discussion in \S5, we adopt a distance of 11 kpc for the two SNRs, HII regions, and both HESS\,J1640$-$465 and HESS\,J1641$-$463.

\section{Mopra observations and data reduction}
For the 7 mm targeted studies, initial Mopra observations towards HESS\,J1640$-$465 and HESS\,J1641$-$463 were taken in April 2012. Four Mopra `On-the-fly' (OTF) 20$'$ by 20$'$ area maps were taken resulting in a 40$'$ by 40$'$ region centred at [$l,b$]=[338$^{\circ}$.26, $-0^{\circ}.072$]. The scan length was 7$''$.6 per cycle time of 2.0 seconds with spacings of 31$''$.2 between each scan row. Each scan consisted of 79 cycles ($\sim$158 seconds). After every 2 scans, a sky reference position was observed for 18 cycles ($\sim$36 seconds) which was used for subtraction in the data reduction process. 3 passes were observed towards each 20$'$ by 20$'$ region in alternating $l$ and $b$ scanning directions. This resulted in $\sim$9 hours of observations per 20$'$ by 20$'$ map. The 7 mm coverage is indicated by the large dashed black box in left panel of Figure \ref{fig:12co(1-0)_int}.

In May 2013 additional observations were taken in a smaller, 12$'$ by 12$'$ region centred at [$l,b$]=[338$^{\circ}$.51,$-0^{\circ}.083$] towards HESS\,J1641$-$463. Similar scan parameters were used as in the 20$'$ by 20$'$ maps over 6 new passes resulting in $\sim$3 times the observation time in this region. The additional observations resulted in greater sensitivity and a lower $T_{\text{RMS}}$ by a factor of $\sim$1.7. This region is indicated by the dashed red box in the left panel of Figure \ref{fig:12co(1-0)_int}.

The Mopra spectrometer, MOPS, was used to target specific molecular line tracers. MOPS is capable of recording in sixteen 4096-channel bands simultaneously whilst in its `zoom' mode as employed here in our 7 mm observations. The list of the targeted molecular transitions and $T_{\text{RMS}}$ levels are displayed in Table \ref{table:mops}. The beam FWHM of Mopra across the 7 mm band varies from 1$'$.37 (31 GHz) to 0$'$.99 (49 GHz), and the velocity resolution of 7 mm zoom-mode data is $\sim$0.2 km/s.

\begin{table}
\caption{The Mopra Spectrometer (MOPS) set-up for 7 mm observations. Displayed are the targeted molecular lines, targeted frequencies, whether the line was detected in our observations and the achieved mapping $T_{\text{RMS}}$.}
\label{table:mops}
\begin{tabular}{|l|c|c|c|}
\hline 
 Molecular line & Frequency & Detection & $T_{\text{RMS}}$ $^{\dagger}$ \\
  & (GHz) & & (K/channel)\\
\hline 
 $^{30}$SiO(J=1-0, v=0) & 42.373365 & - & 0.04\\ 
 SiO(J=1-0, v=3) & 42.519373 & - & 0.04\\ 
 SiO(J=1-0, v=2) & 42.820582 & - & 0.04\\ 
 $^{29}$SiO(J=1-0, v=0) & 42.879922 & - & 0.04\\ 
 SiO(J=1-0, v=1) & 43.122079 & - & 0.04\\ 
 SiO(J=1-0, v=0) & 43.423864 & Yes & 0.04\\ 
 CH$_{3}$OH-I & 44.069476 & Yes & 0.04\\ 
 HC$_{7}$N(J=40-39) & 45.119064 & - & 0.04\\ 
 HC$_{5}$N(J=17-16) & 45.264750 & - & 0.04\\ 
 HC$_{3}$N(J=5-4, F=4-3) & 45.490264 & Yes & 0.05\\ 
 $^{13}$CS(J=1-0) & 46.247580 & - & 0.05\\ 
 HC$_{5}$N(J=16-15) & 47.927275 & - & 0.05\\ 
 C$^{34}$S(J=1-0) & 48.206946 & Yes & 0.06\\ 
 OCS(J=4-3) & 48.651604 & - & 0.06\\
 CS(J=1-0) & 48.990957 & Yes & 0.06\\ 
\hline
\end{tabular}

$^{\dagger}$ Map $T_{\text{RMS}}$ values are for the smaller 12 $'$ by 12 $'$ region described in the text. This is where detections in all of the 7 mm lines were made except for CS(J=1-0). For the detections in CS(J=1$-$0) outside the 12$'$ by 12$'$ region, the $T_{\text{RMS}}$ value was $\sim$ 0.1.
\end{table}

The CO(1-0) line emission data is from the Mopra Southern Galactic Plane CO Survey \citep{2013PASA...30...44B, 2015PASA...32...20B}. This is a survey in the $^{12}$CO, $^{13}$CO and C$^{18}$O $J$ = 1-0 lines over the $l$=305$^{\circ}$-345$^{\circ}$, b=$\pm$0$^{\circ}$.5 region of the Galaxy. The beam FWHM and spectral resolutions of the survey are 0$'$.6 and 0.1 km/s respectively. Full details about the observational parameters used in this survey can be found within the aforementioned papers.

OTF mapping data was reduced and analysed using ATNF analysis software, \textsf{Livedata}, \textsf{Gridzilla}, and \textsf{Miriad}, as well as custom \textsf{IDL} routines.

\textsf{Livedata} was used to calibrate each scan row/column data against a sky reference position and to apply a polynomial baseline-subtraction. \textsf{Gridzilla} was used to re-grid and combine the data from multiple mapping scans into individual three-dimensional data cubes. \textsf{Miriad} and custom \textsf{IDL} routines were used to generate integrated velocity, peak velocity, and position-velocity images from the data cubes.

\section{Gas parameter calculations (spectral line analysis)}
\label{sec:calc}
To investigate the gas distribution towards HESS\,J1640$-$465 and HESS\,J1641$-$463, we calculated mass and density parameters using CO(1-0), CS(1-0) and HI data.
Using the custom \textsf{IDL} routine \textsf{domom}, we produced integrated intensity maps of the different molecular lines. The average column density of molecular hydrogen, $\overline{N_{H_{2}}}$, was calculated from these maps following the corresponding methods outlined in the subsequent sections. The mass of gas in a region, $M$, is then estimated via the relation:
\begin{equation}
M = 2m_{H}\overline{N_{H_{2}}}A
\end{equation}

where $A$ is the cross-sectional area of the region and $m_{H}$ is the mass of a hydrogen atom. From this, the average number density of the region $\overline{n}$ was estimated assuming that the thickness of the region (along the line of sight) had the same value as the height and width. Note that before the intensity values from the velocity-integrated maps could be used to find the column density, they first had to be scaled by a correction factor to account for the beam efficiencies of Mopra at different frequencies. The Mopra extended beam efficiency at 115 GHz (CO(1-0) lines) is $\eta_{\text{XB}} = 0.55$ \citep{2005PASA...22...62L}, while in the 7 mm band at 49 GHz (CS(1-0) lines) $\eta_{\text{XB}} = 0.56$ \citep{2010PASA...27..321U}.

\subsection{CO}
\label{sec:CO_calc}
In this work, to convert brightness temperature to column density, we have adopted the value of the CO(1-0) \mbox{X-factor} to be $X_{\text{CO(1-0)}} \sim 1.5 \times 10^{20}$ cm$^{-2}$(K km/s)$^{-1}$ \citep{2004A&A...422L..47S}. This allows us to calculate the average H$_{2}$ column density in a region, $\overline{N_{H_{2}}} = X_{\text{CO(1-0)}} W_{\text{CO(1-0)}}$, where $W_{\text{CO(1-0)}}$ is the measured $^{12}$CO(1-0) intensity.

The optical depth of the $^{12}$CO line, $\tau_{12}$, was calculated by comparing $^{12}$CO and $^{13}$CO line emission. Following \cite{2013PASA...30...44B}, in the limit where the $^{12}$CO line is optically thick and the $^{13}$CO line is optically thin, $\tau_{12}$ is given by:
\begin{equation}
\tau_{12} = \dfrac{X_{12/13}}{R_{12/13}}
\end{equation}

where $R_{12/13}$ is the ratio of the brightness temperature of the $^{12}$CO and $^{13}$CO lines and $X_{12/13}$ = [$^{12}$C/$^{13}$C] is the isotope abundance ratio. This abundance ratio was determined via $X_{12/13} = 5.5R + 24.2$ where $R$ is the galactocentric radius in kpc \citep{1982A&A...109..344H}.

\subsection{CS}
\label{subsection:CS}
The CS(J=1) column density was calculated using Equation\,9 from \cite{1999ApJ...517..209G}. This equation expresses the upper level column density in terms of the observed integrated line intensity.  The optical depth term required in this equation was determined from the \mbox{CS(1-0)-C$^{34}$S(1-0)} intensity ratio in regions where \mbox{C$^{34}$S(1-0)} was detected. We adopted the elemental abundance ratio of 22.5 for \mbox{[CS]/[C$^{34}$S]} and calculate the optical depth following Equation 1 of \cite{1994A&A...288..601Z}.

Assuming local thermodynamic equilibrium (LTE) at $T_{\text{rot}} \sim 10$ K, the total column density of CS is a factor $\sim 3.5$ times that of the CS(J=1) column density. This temperature assumption introduces a small systematic error into our CS(1-0) column density estimates. A factor of 0.7-1.2 error would be associated with a temperature variation between 5-15 K. We assume a molecular abundance of CS to molecular hydrogen to be $\sim10^{-9}$ \citep{1980ApJ...240...65F}. 

\subsection{HI}
\label{sec:HI}
The Southern Galactic Plane Survey (SGPS) \citep{2005ApJS..158..178M} provided HI data towards HESS\,J1640$-$465 and HESS\,J1641$-$463. Strong absorption features due to continuum sources are seen in data corresponding to the HII regions G338.4+0.0, G338.45+0.06, and G338.39+0.16. Where HI emission features are present we calculate the column density using an HI X-factor, $X_{\text{HI}} = 1.823 \times 10^{18}$ cm$^{-2}$(K kms$^{-1}$)$^{-1}$ \citep{1990ARA&A..28..215D}.

\section{Results}
\label{sec:results}

\begin{figure*}
  \centering
    \includegraphics[width=0.9\textwidth]{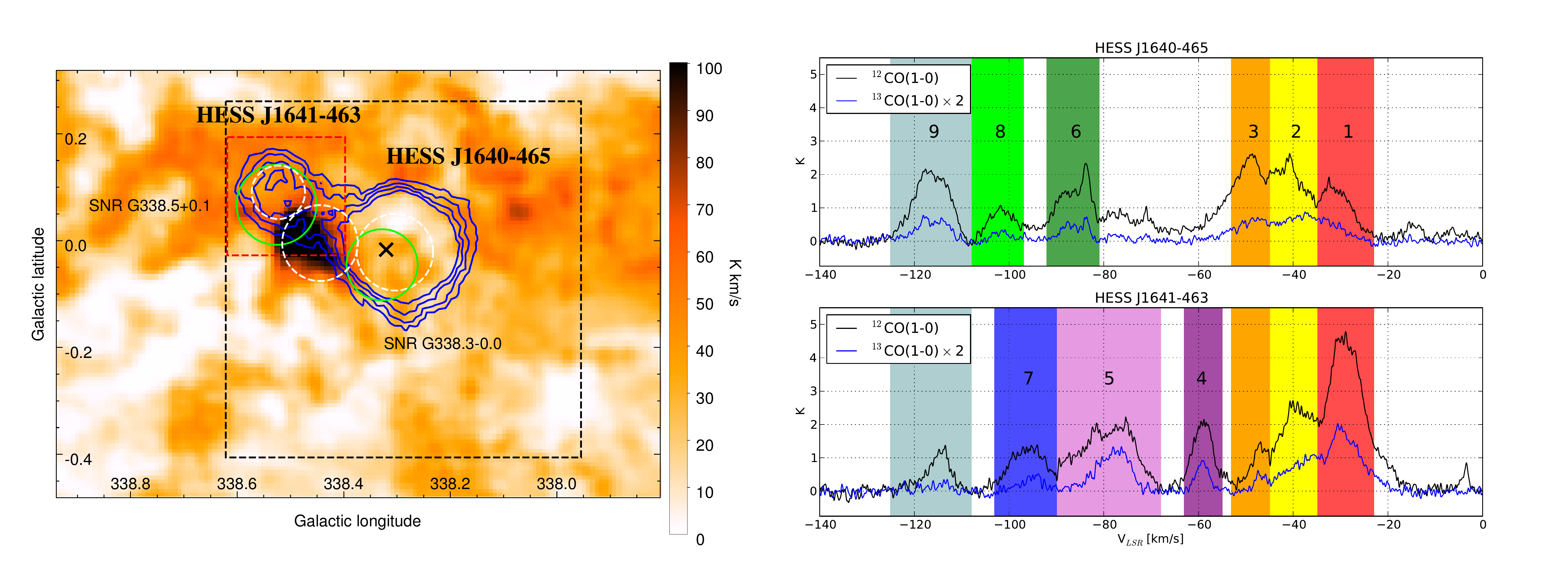}
    \rule{40em}{0.5pt}
  	\caption[what]{\emph{Left:} Mopra $^{12}$CO(1-0) image [K km/s] integrated between -53 and -23 km/s towards HESS\,1641$-$463 and HESS\,1640$-$465. Blue contours are H.E.S.S. significance contours towards the TeV sources at the $5\sigma$, $6\sigma$, $7\sigma$ and $8\sigma$ levels at E $>$ 4 TeV \citep{2014ApJ...794L...1A}. Dashed black and red boxes are the extent of the 7 mm observations and region of additional observations as discussed in text.  Solid green circles are the positions of the labelled SNRs. The black X is the position of PSR\,J1640$-$4631. The 3 dashed white circles are the regions from which CO spectra were extracted, as discussed in text. \emph{Right:} Solid black and blue lines are the average $^{12}$CO(1-0) and $^{13}$CO(1-0) emission spectra respectively within the rightmost and leftmost dashed white circles towards the TeV sources. $^{13}$CO scaled by factor of 2 for clarity. Velocity integration intervals used in Figure \ref{fig:co_big_moasic} are numbered and indicated by the shaded boxes.}
  \label{fig:12co(1-0)_int}
\end{figure*}

\begin{figure*}
  \centering
    \includegraphics[width=0.83\textwidth]{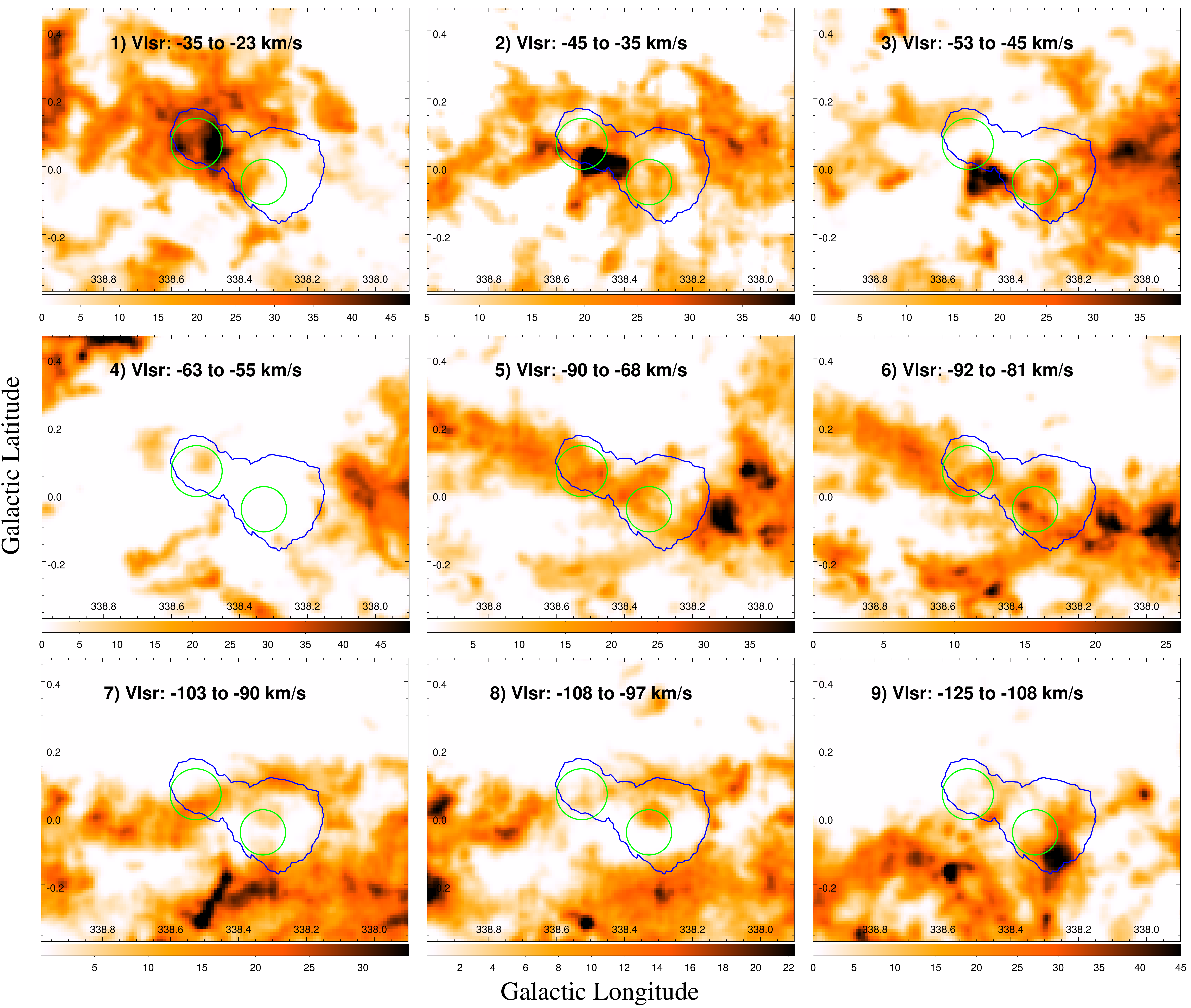}
    \rule{40em}{0.5pt}
  	\caption[what]{Integrated $^{12}$CO(1-0) emission images [K km/s] over indicated velocity intervals. Single blue $5\sigma$ significance H.E.S.S. contour used for clarity. The position and extent of SNR G338.5+0.1 and SNR G338.3-0.0 are indicated by the left and right solid green circles respectively in each panel.}
  \label{fig:co_big_moasic}
\end{figure*}

Overall, the CO transitions reveal a distribution of gas along the line-of-sight towards the TeV sources. Significant detections made in CS(1-0) transitions reveal dense molecular cores within the gas distribution. We also note that detections were made in the SiO(1-0), CH$_{3}$OH\,(I) and HC$_3$N(5-4, F=4-3) transitions towards the dense cores. We discuss these detections in more detail below.

\subsection{CO(1-0) emission}
$^{12}$CO and $^{13}$CO line emission data from the Mopra survey was studied towards HESS\,J1640$-$465 and HESS\,J1641$-$463. $^{12}$CO(1-0) is the standard tracer for diffuse molecular hydrogen gas, while the $^{13}$CO(1-0) line is generally optically thin with $^{13}$CO being approximately 50 times less abundant than $^{12}$CO. Detections in both isotopologues lines were made towards the TeV sources, as well as in other adjacent regions.

\subsection*{CO(1-0) emission towards HESS\,J1640$-$465 and \mbox{HESS\,J1641$-$463}}

A substantial amount of CO(1-0) emission appears to be overlapping the line of sight towards HESS\,J1640$-$465 and HESS\,J1641$-$463. The left panel in Figure \ref{fig:12co(1-0)_int} is an integrated emission image of $^{12}$CO(1-0) data from the Mopra survey between $-53$ and $-23$ km/s towards the two TeV sources. The right panel of Figure \ref{fig:12co(1-0)_int} displays the average $^{12}$CO(1-0) and $^{13}$CO(1-0) emission spectra of the regions corresponding to the reported intrinsic Gaussian size of HESS\,J1640$-$465 and the maximum Gaussian extent of HESS\,J1641$-$463 indicated in the left panel by rightmost and leftmost white dashed circles respectively. The central white dashed circle in the left panel of Figure \ref{fig:12co(1-0)_int} indicates the location of intense CO(1-0) emission seen towards a region that bridges HESS\,J1640$-$465 and HESS\,J1641$-$463, and is discussed further in a later part of this section.

Multiple broad emission components are seen in the CO(1-0) spectra towards HESS\,J1640$-$465 and HESS\,J1641$-$463 in the right panel of Figure \ref{fig:12co(1-0)_int} between $-140$ and 0 km/s. The velocity positions of these components are indicated by the shaded rectangles in the right panel of Figure \ref{fig:12co(1-0)_int} and are labelled as indicated. Figure \ref{fig:co_big_moasic} shows panels of the integrated $^{12}$CO(1-0) emission over said velocity intervals. 

Components 1 2 and 3 (red, yellow and orange shaded boxes) appear both towards HESS\,J1640$-$465 and HESS\,J1641$-$463 and are centred at $\sim$ $-30$, $-40$ and $-50$ km/s respectively. The gas traced in these components are the most likely candidates to be associated with the HII complex, which includes  G338.4+0.0 and G338.45+0.06 with V$_{\text{LSR}}$ $\sim$ $-30$ to $-40$ km/s. This motivates the integration range used in the left panel of Figure \ref{fig:12co(1-0)_int} which corresponds to the velocity space spanned by these three components.

In component 1 ($-35$ to $-23$ km/s), the $^{12}$CO(1$-$0) emission is very prominent in the spectrum towards HESS\,J1641$-$463. The corresponding integrated intensity image shows a molecular cloud positionally coincident with HESS\,J1641$-$463 that extends spatially to at least the Galactic-west and Galactic-north-western parts of SNR G338.3-0.0.

Gas is seen overlapping both TeV sources, as well as all around these sources, in the integrated image for component 2 ($-45$ to $-35$ km/s). Additionally, intense CO emission appears in the region between the TeV sources. An approximate ring of emission can be made out towards HESS\,J1641$-$463 and is discussed in a later section.

In the integrated image for component 3 ($-53$ to $-45$ km/s), the gas overlapping HESS\,J1640$-$465 appears to be connected to a cloud complex to the Galactic-west. Less gas appears to be directly overlapping HESS\,J1641$-$463 and the intense emission between the two TeV sources appears more towards the Galactic-south than in component 2.

The broad features in components 1, 2, and 3 in both $^{12}$CO and $^{13}$CO spectra appear to overlap each other. Thus it is difficult to say with certainty if these are physically connected structures. As such the mass and density parameters for these features were calculated individually. The spectrum in these components were fit with a multi-Gaussian function, and the individual Gaussian functions were used to calculate mass and density parameters. The parameters of the fitted Gaussian functions and the calculated properties of the diffuse H$_{2}$ gas are displayed in Table \ref{table:co_mass_param}.

Emission in component 4 ($-63$ to $-55$ km/s) is seen only in the region towards HESS\,J1641$-$463. A small molecular cloud appears overlapping the Galactic-north upper half of TeV source.

Component 5 ($-90$ to $-68$ km/s) and component 6 ($-92$ to $-81$ km/s) both include emission in a long band of gas that passes through both TeV sources. Emission in component 6 has an additional tail end that extends to the Galactic-south-east of HESS\,J1640$-$465.

Component 7 ($-103$ to $-90$ km/s) has an arm like structure of emission that overlaps through HESS\,J1641$-$463 while a minor amount of wispy gas is seen in component 8 ($-108$ to $-97$ km/s) in the Galactic-northern region of HESS\,J1640$-$465.

Component 9 ($-125$ to $-108$ km/s) includes features in the gas that overlap much of HESS\,J1640$-$465, and appears to be connected to a gas structure immediately to the Galactic-south.

As mentioned in \S\ref{sec:distance}, the HII complex and both SNRs have been established in literature to be at the far distance, with associated kinematic velocities of $\sim -40$ to $-30$ km/s. This corresponds to a distance of $\sim$ 11 to 12 kpc along the line-of-sight (using the rotation curve from \citealt{2007A&A...468..993K}). The gas traced in components 1, 2, and 3 are the only candidates for association to the HII complex and SNRs from kinematic distance considerations, as the v$_{\text{LSR}}$ of other gas components along the line-of-sight will not yield a far distance solution of $\sim$ 11 to 12 kpc.

It is possible that not all gas traced in components 1, 2, and 3 are located at the far distance, as contamination from molecular material located at the near solution may occur. However, the likely need for molecular gas to support the HII complex suggests that a significant fraction of the CO emission in components 1, 2, and 3 traces associated gas located at the far distance.

\begin{table*}
  \centering
  \caption{ $^{12}$CO(1-0) line parameters, and the corresponding calculated gas parameters, from the apertures as indicated in Figure \ref{fig:12co(1-0)_int}. The line-of-sight velocity, v$_{\text{LSR}}$, line-width (full-width-half-maximum), $\bigtriangleup\text{v}_{\text{FWHM}}$, and peak intensity, T$_{\text{peak}}$, were found by fitting Gaussian functions to the \mbox{$^{12}$CO(1-0)} spectra. The optical depth was calculated by comparing the $^{12}$CO and $^{13}$CO line emission following \S\ref{sec:CO_calc}. Masses and density have been scaled to account for an additional 20$\%$ He component. }
  \begin{tabular}{|c|c|c|c|c|c|c|c|c|c|}
  \hline 
  Component & Region & $\text{v}_{\text{LSR}}$ & Distance$^{a}$ & $\bigtriangleup\text{v}_{\text{FWHM}}$ & Peak & Optical & $\overline{N_{H_{2}}}$ $^{b}$ & Mass $^{b}$ & $\overline{n}$ $^{b}$ \\ 
  • &   & (km/s) & (kpc) & (km/s) & (K) & depth & ($10^{21}$ cm$^{-2}$) & (M$_{\odot}\times10^{4}$) & ($10^{2}$ cm$^{-3}$) \\
  \hline 
  1 & HESS\,J1640$-$465 & -31.6 $\pm$ 1.1 & 11.9 & 3.9 $\pm$ 0.1 & 1.6 $\pm$ 0.1 & 10.5 & 4.3  & 6.8  & 2.0 \\

    & HESS\,J1641$-$463 & -29.0 $\pm$ 0.1 & 11.9 & 4.1 $\pm$ 0.1 & 4.5 $\pm$ 0.1 & 11.7 & 12.8 & 9.7  & 8.4 \\
  
    & Bridge            & -28.1 $\pm$ 0.1 & 11.9 & 5.2 $\pm$ 0.1 & 3.5 $\pm$ 0.1 & 13.1 & 12.5 & 19.0 & 5.8 \\ 
  \hline
  2 & HESS\,J1640$-$465 & -40.7 $\pm$ 0.1 & 11.2 & 2.6 $\pm$ 0.1 & 1.9 $\pm$ 0.1 & 8.5  & 3.4  & 4.7  & 1.6 \\ 

    & HESS\,J1641$-$463 & -39.8 $\pm$ 0.1 & 11.2 & 2.7 $\pm$ 0.1 & 2.5 $\pm$ 0.1 & 10.3 & 4.6  & 3.1  & 3.2 \\
  
    & Bridge            & -40.5 $\pm$ 0.1 & 11.2 & 2.4 $\pm$ 0.1 & 4.4 $\pm$ 0.1 & 11.1 & 7.4  & 9.9  & 3.6 \\ 
  \hline
  3 & HESS\,J1640$-$465 & -49.2 $\pm$ 0.1 & 10.8 & 4.2 $\pm$ 0.1 & 2.4 $\pm$ 0.1 & 6.9  & 6.9  & 8.8  & 3.5 \\ 

    & HESS\,J1641$-$463 & -47.0 $\pm$ 0.2 & 10.8 & 2.3 $\pm$ 0.2 & 1.2 $\pm$ 0.1  & 11.1 & 1.8  & 1.1  & 1.3 \\ 

    & Bridge            & -48.3 $\pm$ 0.1 & 10.8 & 4.2 $\pm$ 0.1 & 3.9 $\pm$ 0.1 & 9.3  & 11.0 & 13.9 & 5.6 \\ 

  \hline
\end{tabular}

\begin{flushleft}
$^{a}$ Assumed distances, $d_{0}$, used for mass and density calculations are derived from the Galactic rotation curve presented in \cite{2007A&A...468..993K}. However, these values are easily scaled for an arbitrary distance, $d$, by multiplying by $(d/d_{0})^{2}$ and $(d/d_{0})^{-1}$ for mass and density respectively. \\
$^{b}$ The error in the calculated physical parameters are dominated by the statistical uncertainties associated with the CO to H$_{2}$ conversion factor ($X_{\text{CO(1-0)}}$) and is on the order of 30$\%$ \citep{2013ARA&A..51..207B}.
\end{flushleft}
\label{table:co_mass_param}
\end{table*}

\subsection*{CO bubble feature seen at $v_{\text{LSR}} \sim$ $-40$ to $-35$ km/s}

\begin{figure}
  \centering
    \includegraphics[width=0.5\textwidth]{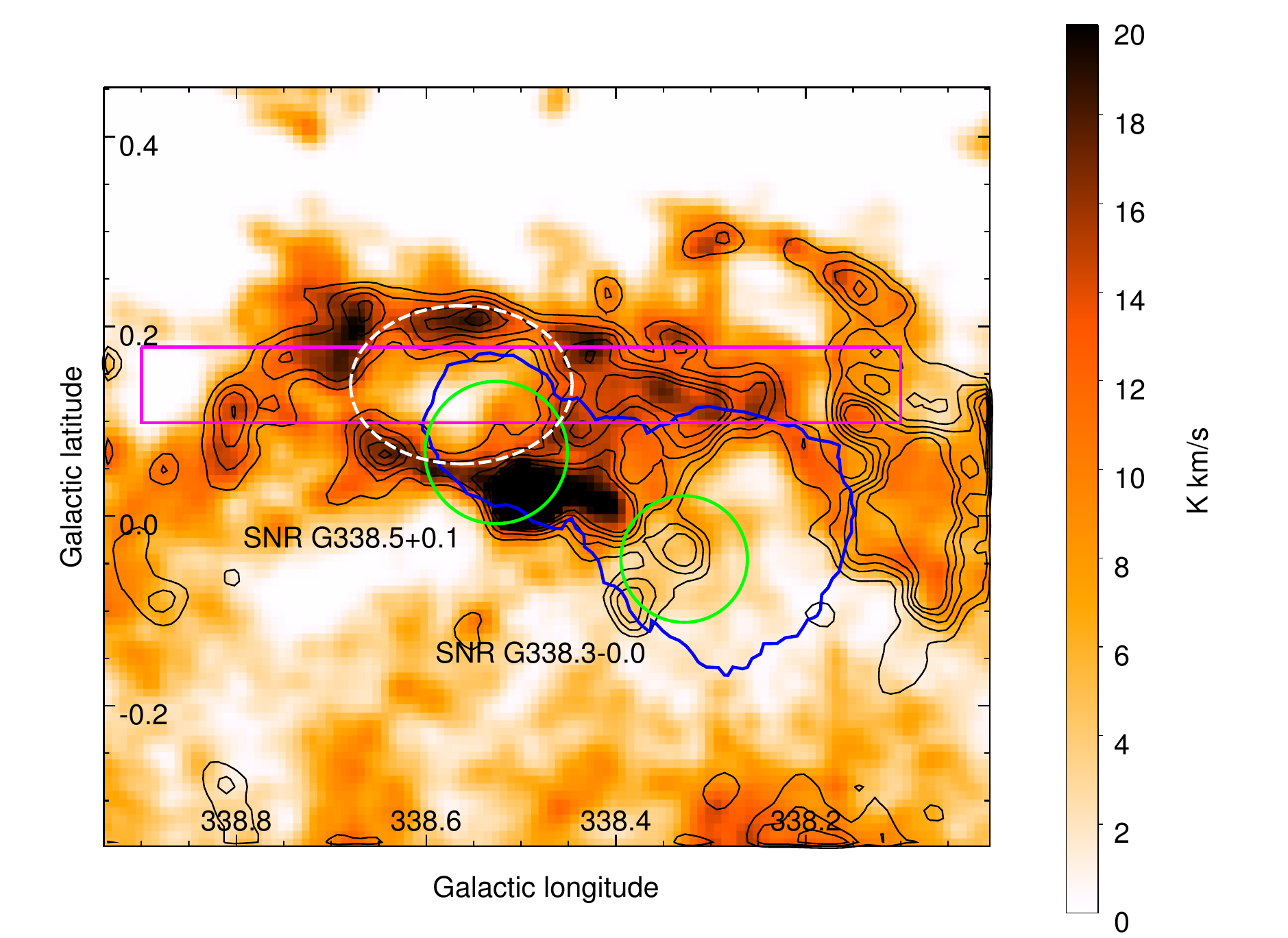}
    \rule{20em}{0.5pt}
  	\caption[what]{Mopra $^{12}$CO(1-0) emission image [K km/s] integrated between $-40$ and $-35$ km/s.  Single solid blue contour is 5$\sigma$ significance from H.E.S.S. observations towards HESS\,1640$-$465 and HESS\,1641$-$463. The white dashed ellipse is the approximate position of a ring feature discussed in the text and the solid magenta box is the integration region for the position-velocity plot shown in Figure \ref{fig:pv_plot_ring}. Overlaid black contours are from Mopra $^{13}$CO(1-0) observations. SNRs are indicated by green circles.}.
  \label{fig:12co(1-0)_ring}
\end{figure}

Figure \ref{fig:12co(1-0)_ring} shows the integrated $^{12}$CO(1-0) emission between $-40$ and $-35$ km/s. Overlaid are black contours indicating integrated $^{13}$CO(1-0) emission in the same velocity interval. Both data show an ellipse-like ring of emission seen approximately positionally coincident with HESS\,J1641$-$463. The location of this ring is indicated by the white dashed ellipse in the figure. The ellipse has semi-major and semi-minor axis lengths of $\sim$ 7 and 5 arcminutes respectively.

\begin{figure}
  \centering
    \includegraphics[width=0.5\textwidth]{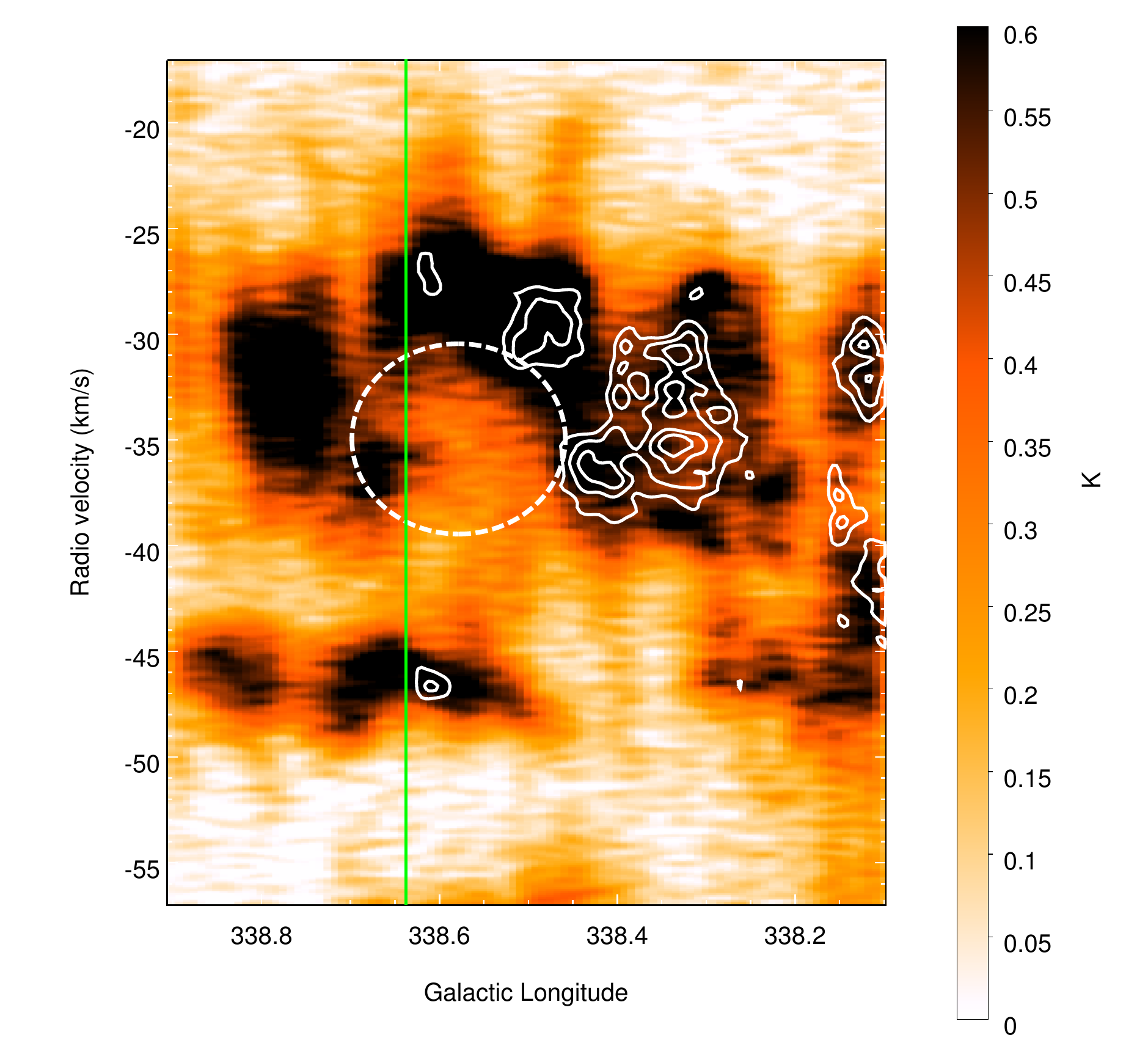}
    \rule{20em}{0.5pt}
  	\caption[]{Position-velocity image [K] (in Galactic longitude) of $^{12}$CO(1-0) emission towards the bubble feature seen towards HESS\,J1641$-$463 in the region indicated in Figure \ref{fig:12co(1-0)_ring}. The approximate position of the cavity discussed in text is indicated by the dashed white ellipse. The solid white contours indicate CS(1-0) emission detected in our 7 mm observations. Note that the extent of the coverage in 7 mm only reaches the vertical green line in longitude.}
  \label{fig:pv_plot_ring}
\end{figure}

Figure \ref{fig:pv_plot_ring} is a position-velocity plot (in longitude) of the $^{12}$CO(1-0) emission in the magenta rectangle region shown in Figure \ref{fig:12co(1-0)_ring}. A cavity is seem in the gas at the \mbox{$\sim$ $-40$} to $-30$ km/s velocity range, the approximate position of which is illustrated by the dashed white ellipse. The overlaid white contours indicate integrated emission in the dense gas tracer CS(1-0) seen in our 7 mm observations in the same velocity interval. Note that the extent of the coverage in 7 mm only partially covers the position-velocity plot. The image suggests that the bubble-like feature may have been blown out from one side of the molecular cloud seen in \mbox{component 1} (velocity range $\sim$ $-35$ to $-23$ km/s) in Figure \ref{fig:co_big_moasic}. It is possible that this bubble has been blown out by the SNR G338.5+0.1 or perhaps the result of the stellar wind from a progenitor star.

From Figure \ref{fig:12co(1-0)_ring}, the thickness of the ring is $\sim$ 3 arcminutes. The kinematic distance along the line-of-sight is $\sim$ 11.4 kpc at $-40$ to $-35$ km/s. Under these assumptions, the column density of the gas enclosed by the ring is \mbox{$\overline{N_{H_{2}}} \sim 4 \times 10^{21}$ cm$^{-2}$}, with a total mass of $\sim$ $8 \times 10^{4}$ M$_{\odot}$. From Figure \ref{fig:pv_plot_ring}, the expansion velocity of the bubble appears to be $\sim5-10$ km/s. The expansion of the bubble would then have a kinetic energy of  $\sim2-8\times10^{49}$ erg.

If this were a wind-blown bubble, an O-type progenitor star with mass \mbox{$\sim27$ M$_{\odot}$} would be able to create it, based on the bubble size of radius $\sim25$ pc at a distance of $\sim$ 11.4 kpc \citep[and references therein]{2013ApJ...769L..16C}. The energy required to produce such a bubble can then be calculated following the model presented by \cite{1999ApJ...511..798C}, and is $\sim 4 \times 10^{49}$ erg.

\subsection*{CO(1-0) emission towards dense `bridge' between \mbox{HESS\,J1640$-$465} and \mbox{HESS\,J1641$-$463}}
\begin{figure}
  \centering
    \includegraphics[width=0.5\textwidth]{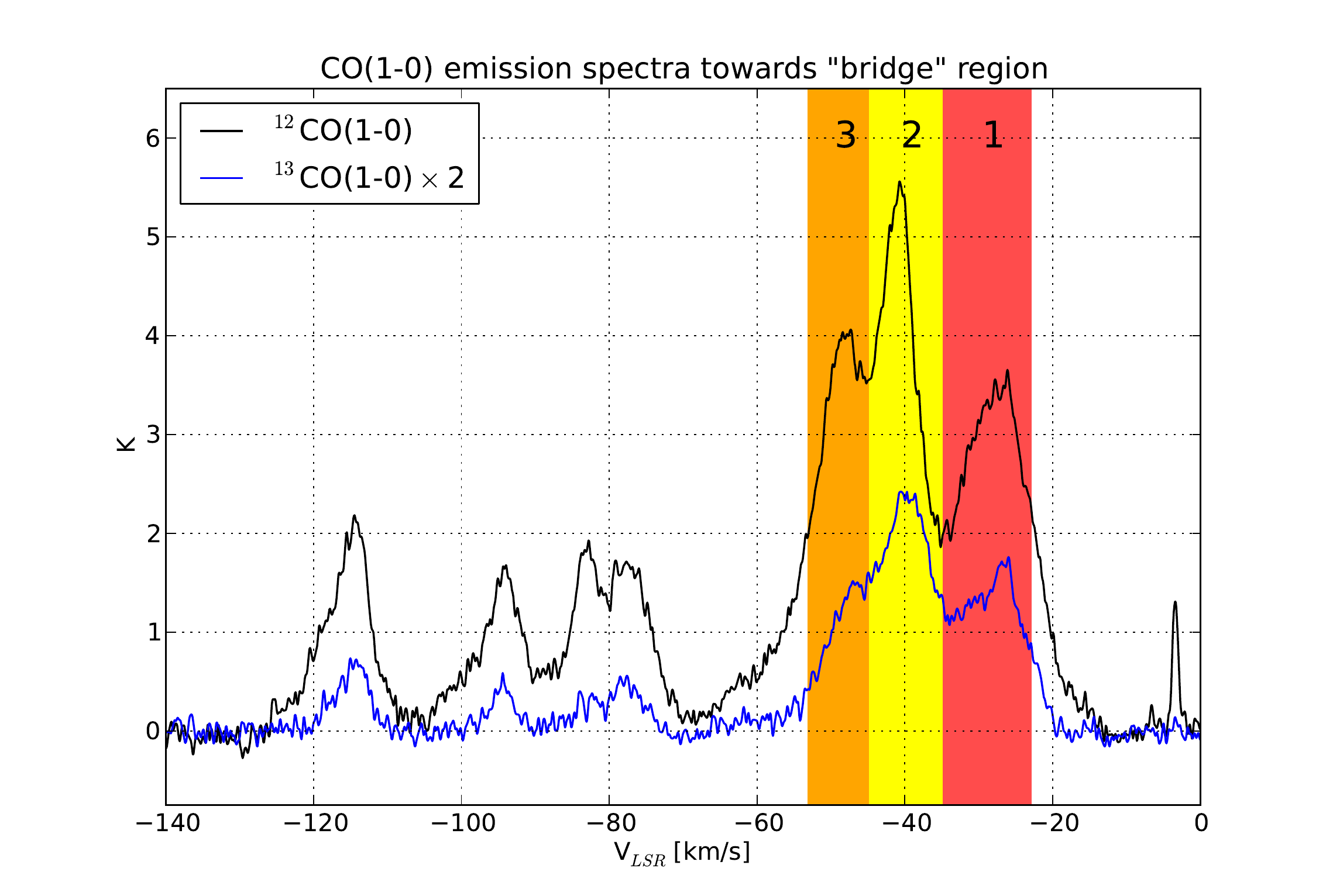}
    \rule{20em}{0.5pt}
  	\caption[what]{$^{12}$CO(1-0) (black) and $^{13}$CO(1-0) (blue) emission spectra towards the ``bridge" between HESS\,J1640$-$465 and HESS\,J1641$-$463 indicated by the central white dashed circle in Figure \ref{fig:12co(1-0)_int}. $^{13}$CO scaled by a factor of 2 for clarity. Velocity ranges for components 1, 2, and 3 are indicated by the shaded boxes.}
  \label{fig:ridge_12CO_spectrum}
\end{figure}

In components 1, 2, and 3 of Figure \ref{fig:co_big_moasic}, we see an area of intense CO emission located towards the Galactic-south part of the region that bridges HESS\,J1640$-$465 and HESS\,J1641$-$463 which appears to span between $\sim$ $-35$ to $-55$ km/s. This region is indicated by the central white dashed circle in Figure \ref{fig:12co(1-0)_int}. Figure \ref{fig:ridge_12CO_spectrum} shows the average $^{12}$CO(1-0) and \mbox{$^{13}$CO(1-0)} emission spectra in this region. 3 components are seen in the $^{12}$CO spectra that match well with components 1, 2, and 3 in Figure \ref{fig:12co(1-0)_int}.
Calculated mass and density parameters for these components are displayed in Table \ref{table:co_mass_param}.

\subsection*{CO(1-0) emission towards the Galactic-west of HESS\,J1640$-$465 and HESS\,J1641$-$463}
\begin{figure}
  \centering
    \includegraphics[width=0.5\textwidth]{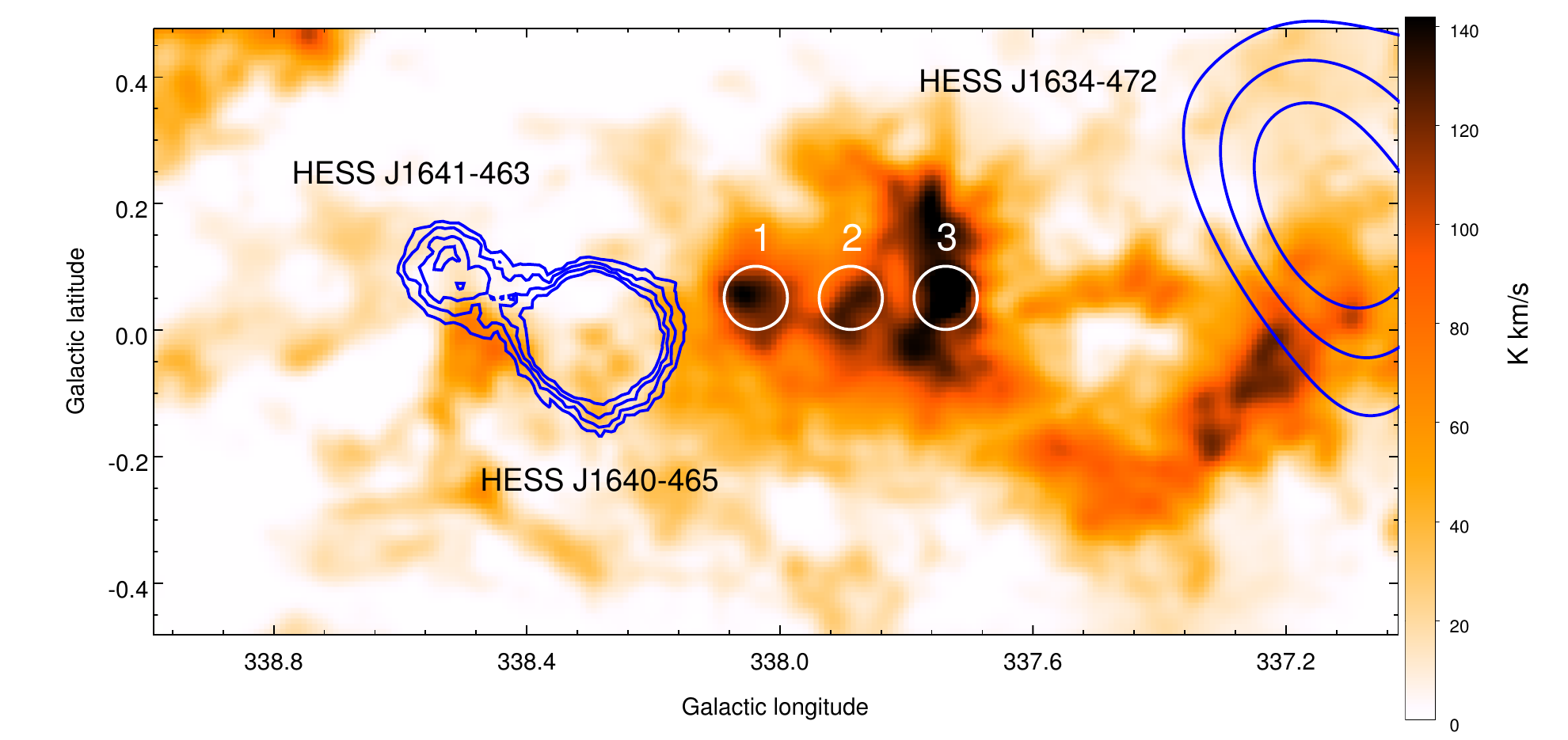}
    
    \includegraphics[width=0.45\textwidth]{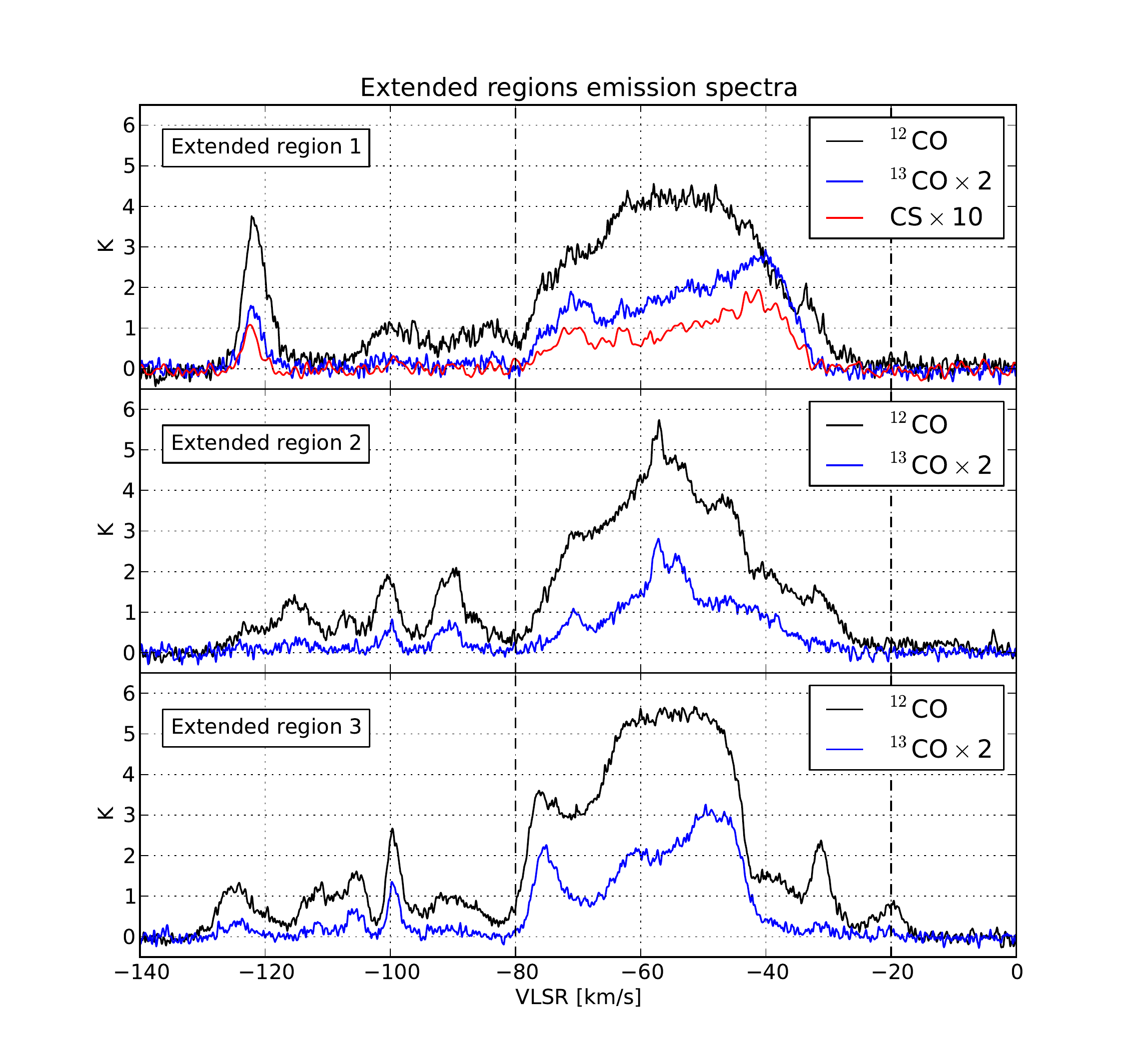}
    \rule{20em}{0.5pt}
  	\caption[what]{\emph{Top:} Integrated $^{12}$CO emission image [K km/s] between $-80$ and $-20$ km/s from $l=339^{\circ}$ to $l=337^{\circ}$. Blue contours indicate the positions of HESS\,J1640$-$465 and HESS\,J1641$-$463 \citep{2014ApJ...794L...1A}. Contours for the TeV source HESS\,J1634$-$472 are also shown in blue for completeness \citep{hess_plane}. \emph{Bottom:} Average $^{12}$CO(1-0) (black) and $^{13}$CO(1-0) (blue) emission spectra in three circular regions indicated above in white. CS(1-0) emission from 7 mm observations is displayed in red for extended region 1. $^{13}$CO and CS(1-0) emission have been scaled by a factor of 2 and 10 respectively for clarity. Vertical dashed lines indicate the integration range used to produce the image in the top panel.}
  \label{fig:co_g338_g337}
\end{figure}

CO(1-0) emission over a very broad ($\sim$ 60 km/s) velocity range can be seen in an extended molecular cloud structure to the Galactic-west of HESS\,J1640$-$465. The top panel of Figure \ref{fig:co_g338_g337} displays an integrated image of the \mbox{$^{12}$CO(1-0)} emission between $-80$ and $-20$ km/s over an extended region from $l=339^{\circ}$ to $l=337^{\circ}$ taken from the Mopra CO survey. Contours of the TeV source HESS\,J1634$-$472 are shown in addition to those of HESS\,J1640$-$465 and HESS\,J1641$-$463 for completeness. Spectra in three representative regions in this extended structure, indicated by white circles in the top panel, are shown in the bottom panel. The average $^{12}$CO(1-0) and $^{13}$CO(1-0) spectra are shown by black and blue lines respectively. Our 7 mm observations had coverage over the region labelled ``1" in the top panel of Figure \ref{fig:co_g338_g337}, and we include the spectrum for CS(1-0) emission in red in the corresponding set of axes. Note that the $^{13}$CO(1-0) and CS(1-0) emission have been scaled by a factor of 2 and 10 respectively for clarity.

Broad emission from $\sim$ $-80$ to $-20$ km/s is seen in these regions which may be due to multiple contributing components. The broadness of this emission make it difficult to place the gas at a distance with any certainty using Galactic rotation curve calculations. The rotation curve from \cite{2007A&A...468..993K} yields a distance estimate of $1.6 - 4.6$ kpc (near solution) and $9.5 - 12.5$ kpc (far solution). The far distance solutions overlap with the estimated distances to HESS\,J1640$-$465 and HESS\,J1641$-$463, so it is important to give consideration to the gas structure as CR target material. This is discussed further in \S \ref{sec:hadronic_sce}.

\subsection{7 mm line emission}
\label{sec:cs}
In our 7 mm observations towards HESS\,J1640$-$465 and HESS\,J1641$-$463, detections were made in the CS(1$-$0), C$^{34}$S(1-0), SiO(J=1-0,v=0), HC$_{3}$N(5-4, F=4-3) and CH$_{3}$OH\,(I) lines.

Dense gas in the region was traced by CS(1-0) and C$^{34}$S(1-0) emission. The CS(1-0) transition has a critical density for emission of $\sim\times10^{5}$ cm$^{−3}$ at a temperature of $\sim$\,10\,K, making it an ideal tracer for probing the deeper and denser inner regions of molecular clouds. SiO emission is usually produced behind shocks moving through molecular clouds \citep{2008A&A...482..809G} from which the \mbox{SiO(J=1-0,v=0)} line can be detected. HC$_{3}$N is often detected in warm molecular clouds and is associated with star forming regions, while the CH$_{3}$OH\,(I) maser generally traces star formation outflows.

\begin{figure*}
  \centering
    \includegraphics[width=0.8\textwidth]{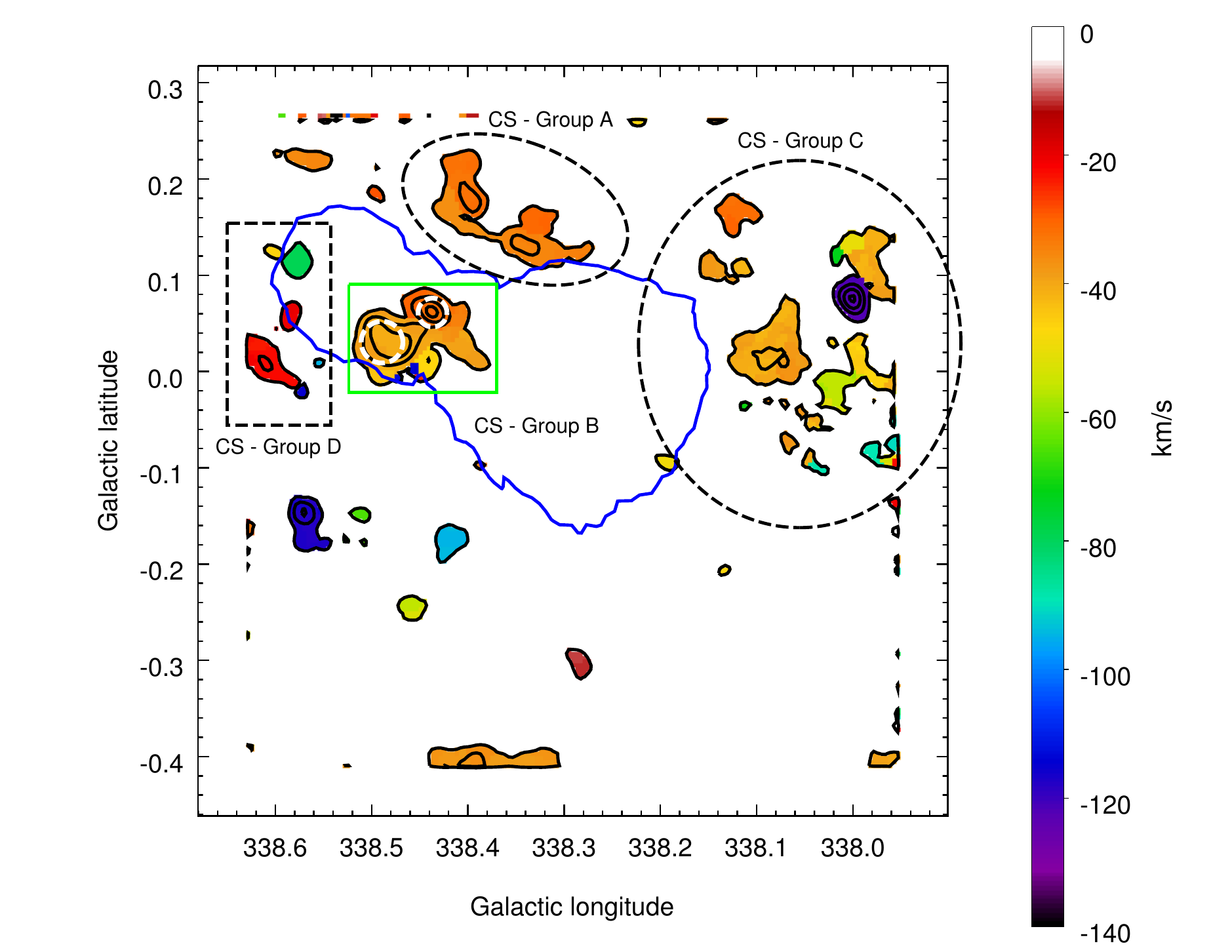}
    \rule{40em}{0.5pt}
  	\caption[what]{Velocity of peak pixel CS(1-0) map [km/s] in the 7 mm observations region. Overlaid solid black contours show the intensity of the peak pixel. H.E.S.S. 5$\sigma$ significance contour is in solid blue. Regions of interest discussed in the text are labelled by black dashed ellipses. The CS - Group B region, however, is outlined by a solid green box and indicates the integration region for the position-velocity plot shown in Figure \ref{fig:cs(1-0)_pv_final}. Dashed white circles indicate apertures used to extract spectra for Bridge Core 1 (left circle) and Bridge Core 2 (right circle) discussed in the text.}
  \label{fig:cs(1-0)_mom-3}
\end{figure*}

The location of the dense gas traced by the CS(1-0) line in our study are displayed via a velocity-of-peak-pixel map in Figure \ref{fig:cs(1-0)_mom-3}. From the figure, we can see that most of strongest \mbox{CS(1-0)} emission occurs at a velocity consistent with components 1, 2, and 3 ($-53$ to $-23$ km/s) in the CO(1-0) data. Several regions of significant CS(1-0) emission present themselves and are roughly grouped together as illustrated in Figure \ref{fig:cs(1-0)_mom-3}. These groups are discussed below together with other detections made in the 7 mm band. The detections of the 7 mm lines aside from CS are shown in Figure \ref{fig:other_7mm} overlaid on a Spitzer 8.0 $\mu$m image of the region.

\begin{figure*}
\captionsetup[subfigure]{labelformat=empty}
     \centering
     \subfloat[][]{\includegraphics[width=0.34\textwidth]{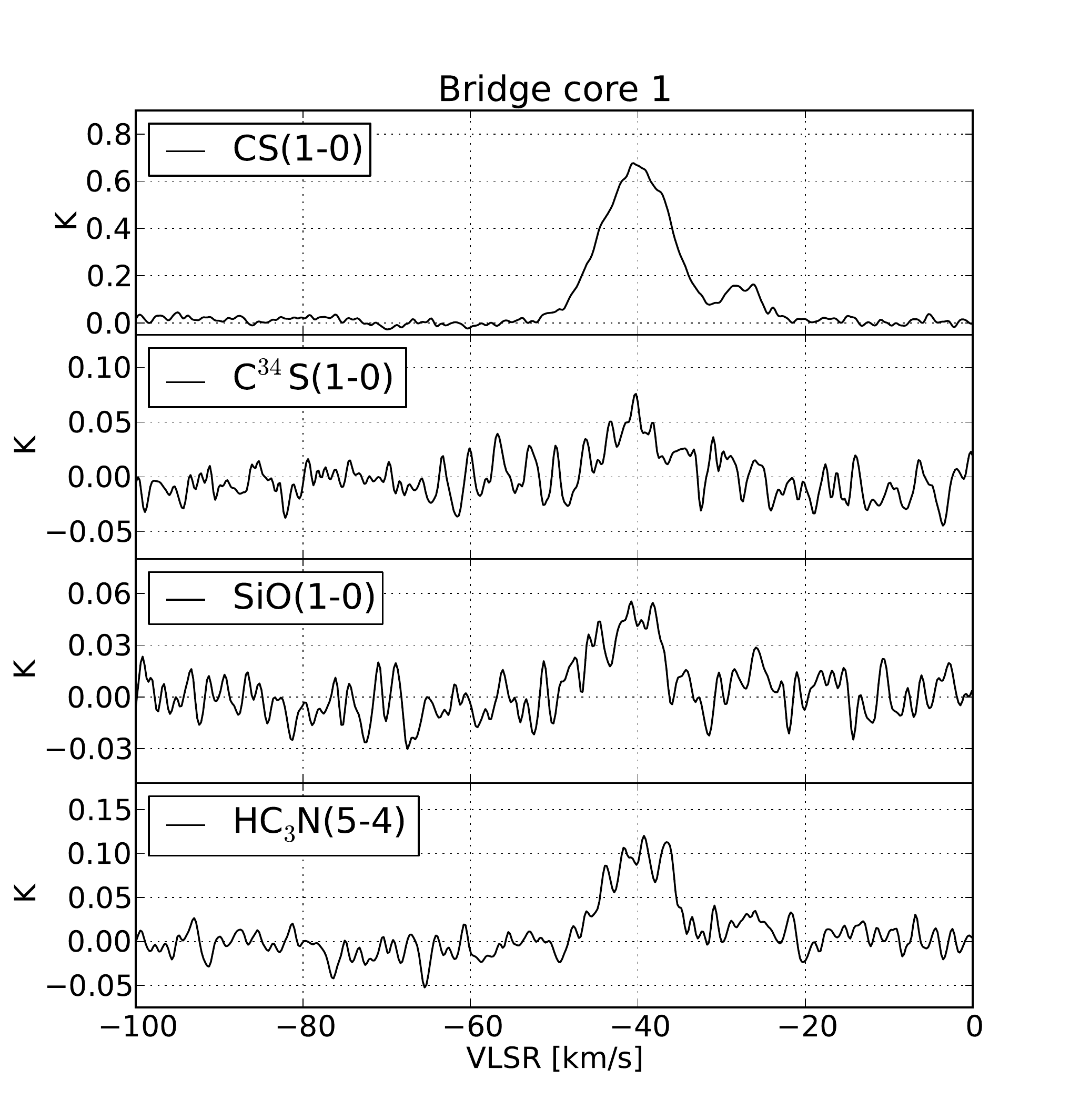}
     \label{core1_cs_spectra}}
     \subfloat[][]{\includegraphics[width=0.34\textwidth]{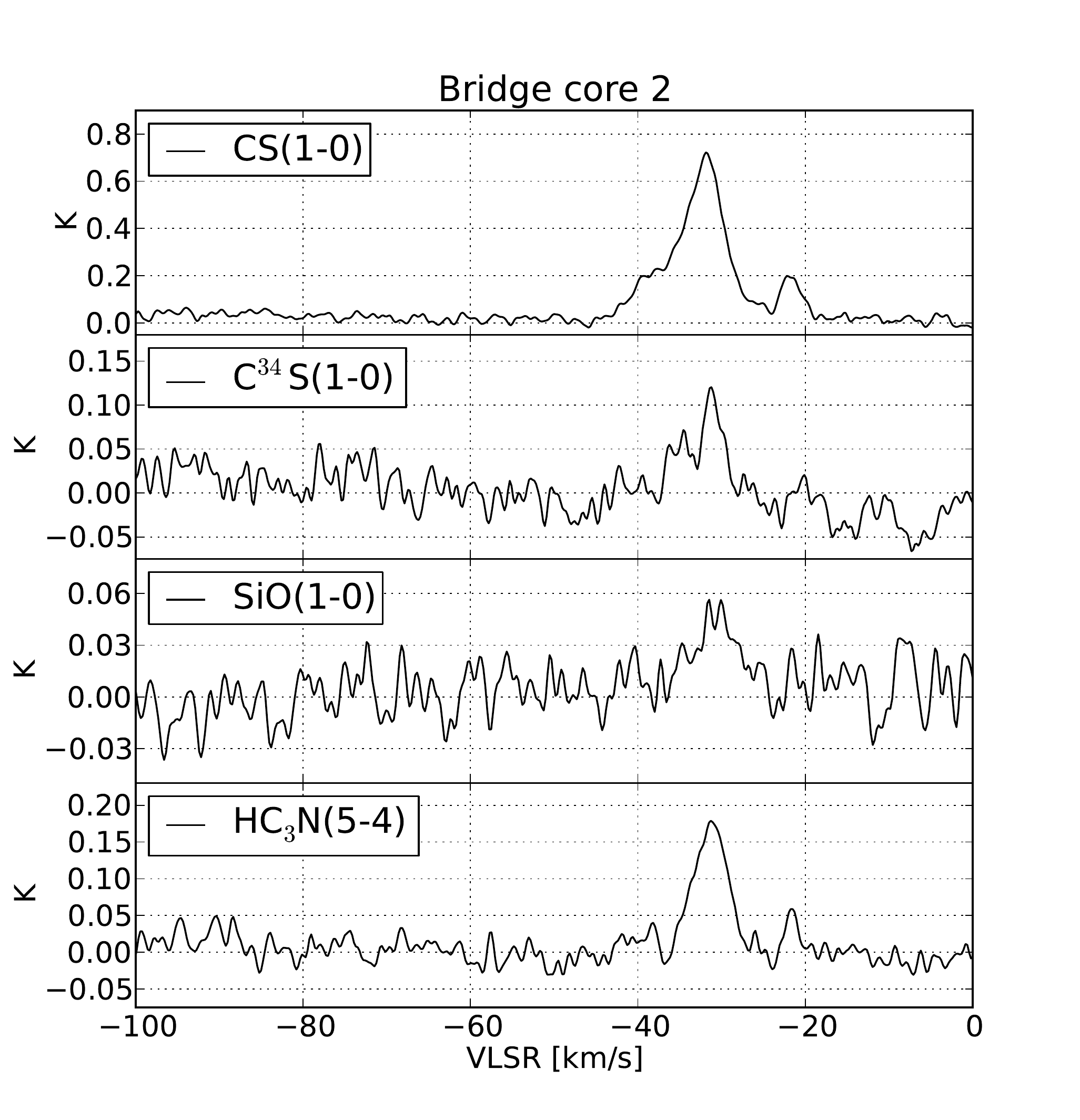}
     \label{core2_cs_spectra}}
     \subfloat[][]{\includegraphics[width=0.34\textwidth]{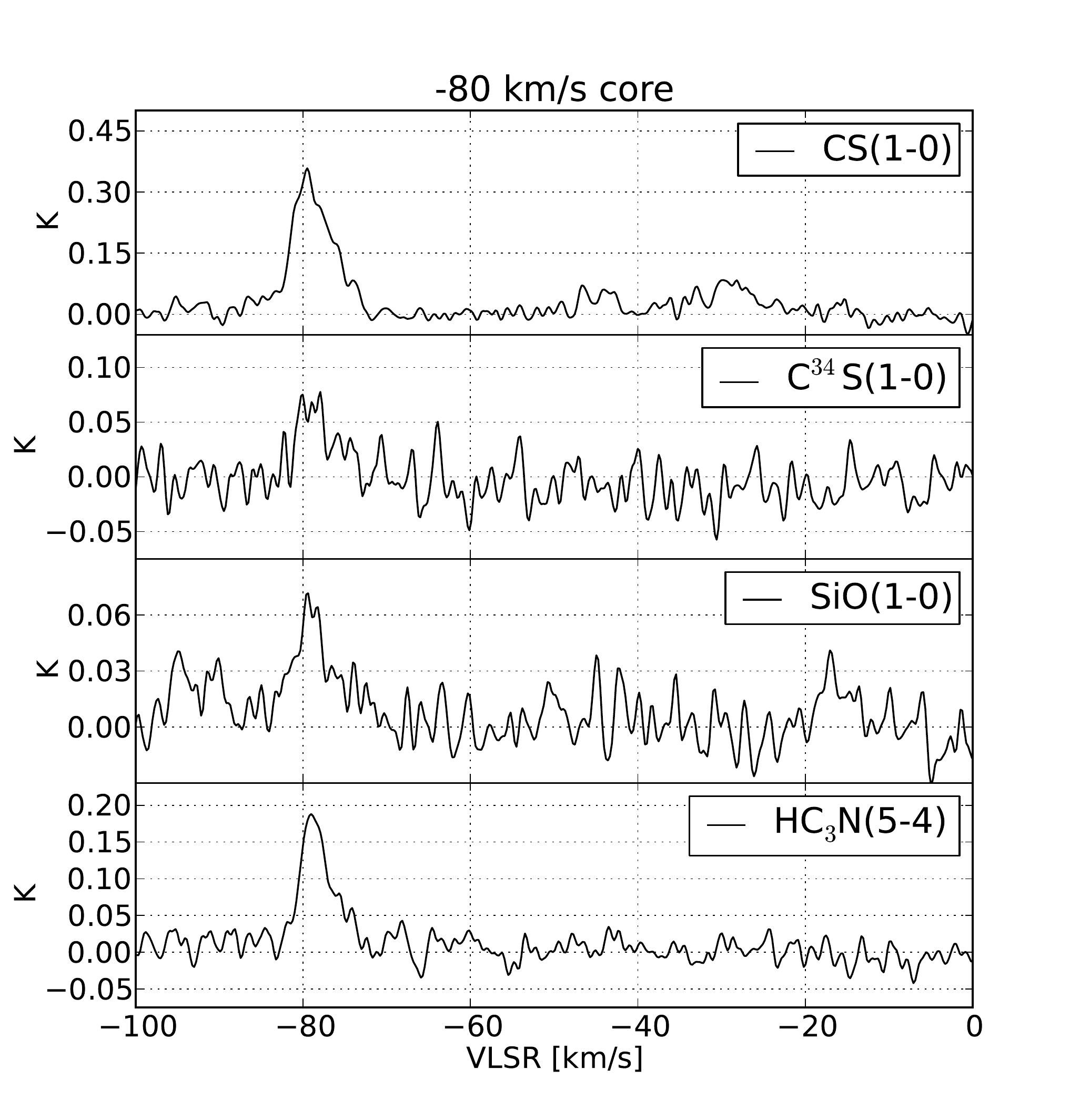}
     \label{-80kms_7mm_spectra}}
     
\caption{Average spectra of detected 7 mm lines within apertures centred at Bridge Core 1 (\textit{Left panel}), Bridge Core 2 (\textit{Centre panel}) and the $-80$ km/s core (\textit{Right panel}).}    
\label{fig:core_cs_spectra}
\end{figure*}

\subsection*{7 mm emission in Group A}
Group A is located slightly to the Galactic-north of HESS\,J1640$-$465 and HESS\,J1641$-$463, and is approximately coincident with a bright HII region G338.39+0.16. CS(1-0) emission in this region appears in the \mbox{$\sim$ $-40$ to $-30$ km/s} velocity range.
The morphology of the emission is extended in nature, forming a slight arc with what appears to be 2 dense clumps.

\subsection*{7 mm emission in  Group B or ``bridge" region}
Group B is the central bridge coincident with the HII complex between HESS\,J1640$-$465 and HESS\,J1641$-$463. A markedly strong amount of CS(1-0) emission is seen here spanning a broad ($\sim$ 20 km/s) velocity space. Group B is also roughly positionally coincident with the intense emission seen in the CO ``bridge" discussed earlier, with emission features at similar velocities.

\begin{figure}
  \centering
    \includegraphics[width=0.48\textwidth]{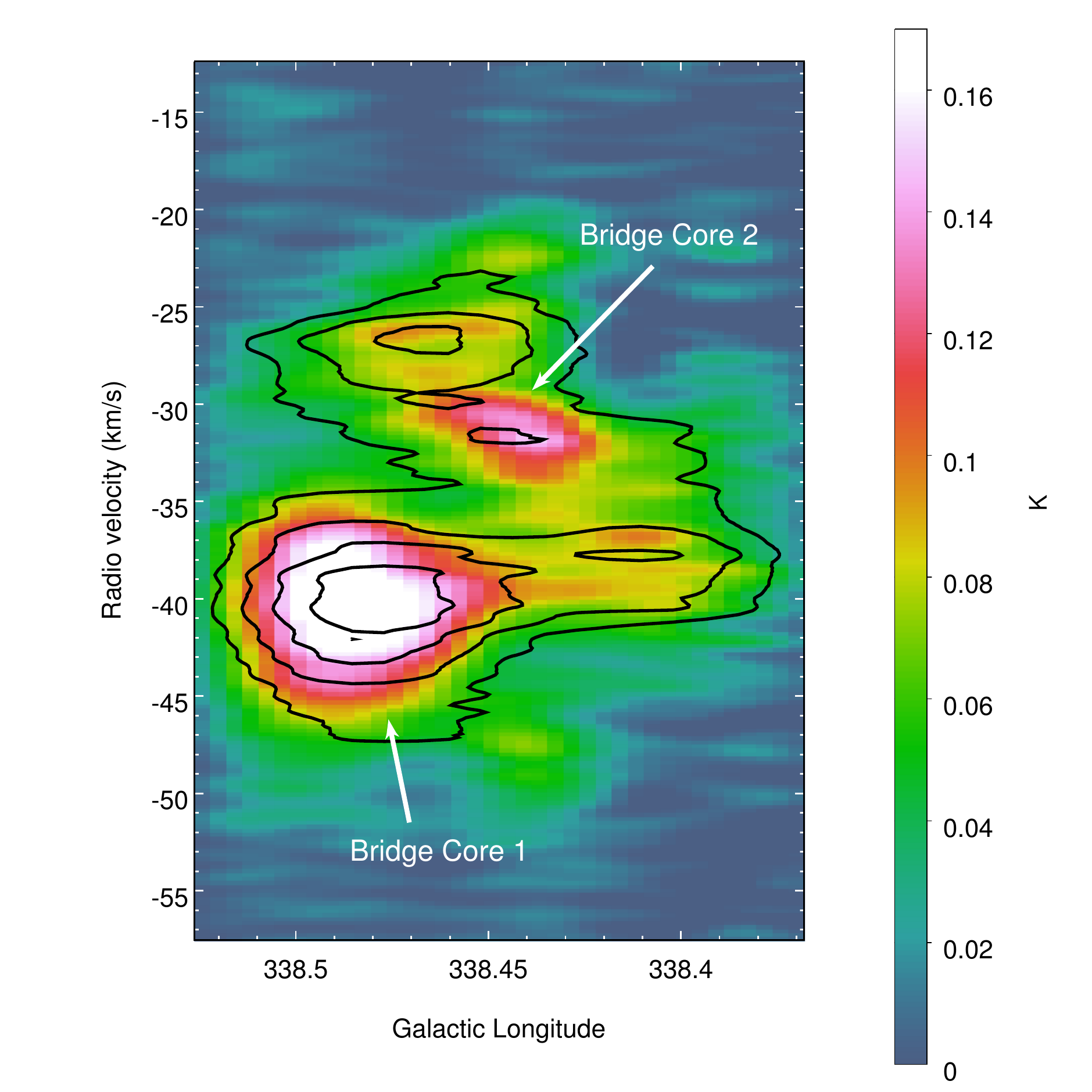}
    \rule{20em}{0.5pt}
  	\caption[what]{Position-velocity image [K] of CS in the region indicated by a solid green box in Figure \ref{fig:cs(1-0)_mom-3}. Locations of Bridge Core\,1 and Bridge Core\,2 are labelled with arrows. Black contours are from Mopra $^{13}$CO(1-0) observations.}
  \label{fig:cs(1-0)_pv_final}
\end{figure}

\begin{figure}
  \centering
    \includegraphics[width=0.47\textwidth]{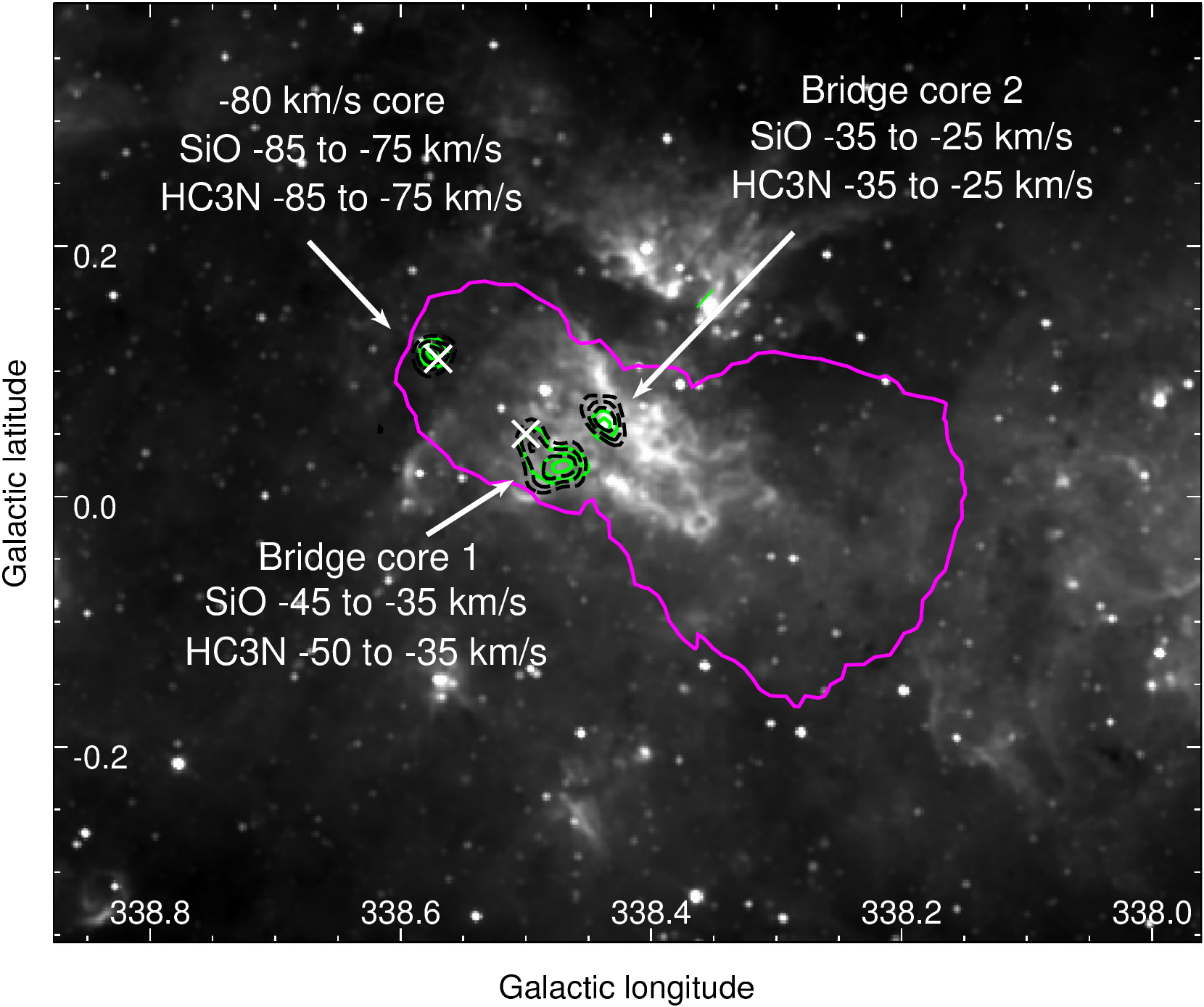}
    \rule{20em}{0.5pt}
  	\caption[what]{Spitzer 8.0 $\mu$m image towards HESS\,J1640$-$465 and HESS\,J1641$-$463. Single magenta contour is 5$\sigma$ significance from H.E.S.S. observations. Green solid contours are integrated SiO(1$-$0) emission. Dashed black contours are integrated \mbox{HC$_{3}$N(5-4, F=4-3)} emission. The velocities over which they are integrated are as labelled. White Xs indicate positions of observed CH$_{3}$OH (I) masers.}
  \label{fig:other_7mm}
\end{figure}

Figure \ref{fig:cs(1-0)_pv_final} is a longitude-velocity image of CS(1-0) data in the region indicated by a solid green box in Figure \ref{fig:cs(1-0)_mom-3}. From this image, there appears to be two separate dense cores with broad CS(1-0) emission in this region separated in velocity, but somewhat overlapping spatially along the line of sight. We label these Bridge Core 1 and Bridge Core 2 as illustrated. These cores appear to be embedded in an extended bridge of emission linking themselves as well as other smaller clumpy features. We note the good correlation between Bridge Core 1 and emission in $^{13}$CO(1-0) (indicated by the solid black contours in Figure \ref{fig:cs(1-0)_pv_final}). Bridge Core 2, however, appears to be offset from the local maximum traced in $^{13}$CO.

Detections in the isotopologue transition C$^{34}$S(1-0), as well as in SiO(1-0) and HC$_{3}$N(5-4, F=4-3)  were made towards core 1 and core 2. The left and centre panels of Figure \ref{fig:core_cs_spectra} displays the emission spectra in these lines towards Bridge Core 1 and Bridge Core 2. The spatial size of the cores was determined by fitting a Gaussian function to the line profile drawn through the centre of each core, and the spectra extracted from circular apertures with sizes equal to the FWHMs (1.3$'$ and 0.9$'$ for Bridge Core 1 and 2 respectively).
We note that all the 7 mm line emission peaks at the same velocity as the intense emission in CS(1-0); $-40$ and $-32$ km/s for Bridge Core 1 and 2 respectively. Table \ref{table:7mm_detections} displays the 7 mm detection parameters and mass estimates calculated from the data in these cores. Mass parameters were calculated following $\S$\ref{subsection:CS} assuming the gas is at a distance of 11 kpc.

SiO(1-0) emission appears positionally coincident with Bridge Core 1 and Bridge Core 2 at the same velocity as that observed in CS(1-0). This suggests that both these cores have been disturbed by a shock passing through. A CH$_{3}$OH\,(I) maser is seen at the Galactic-north-east edge of Bridge Core 1, suggesting the presence of an outflow. Combined with the detection of HC$_{3}$N at the same position and velocity and that Bridge Core 1 and 2 appear to be embedded in the complex of HII regions, the shock is likely to be been caused by recent nearby star formation.

\begin{table*}
  \centering
  \caption{7 mm line parameters extracted from apertures towards Bridge Core 1 and Bridge Core 2 (as shown in Figure\ref{fig:cs(1-0)_mom-3}). The line-of-sight velocity, $\text{v}_{\text{LSR}}$, peak intensity, T$_{\text{peak}}$, and line-width, $\bigtriangleup\text{v}_{\text{FWHM}}$, were found by fitting Gaussian functions to the spectra in \mbox{Figure \ref{fig:core_cs_spectra}}. The optical depth, together with mass and density calculations used the CS(1-0) and C$^{34}$S(1-0) data following \S \ref{subsection:CS}.$^{a}$}
\begin{tabular}{|c|c|c|c|c|c|c|c|c|}
\hline 
Object & Detected lines & $\text{v}_{\text{LSR}}$ & T$_{\text{peak}}$ & $\bigtriangleup\text{v}_{\text{FWHM}}$ & Optical & $\overline{N_{H_{2}}}$ & Mass & $\overline{n}$ \\
   &   & (km/s) & (K) & (km/s) & depth &($\times10^{23}$ cm$^{-2}$) & (M$_{\odot}$) & ($\times 10^{4}$ cm$^{-3}$) \\
\hline 
Bridge Core 1 & CS(1-0) & -40.2 $\pm$ 0.1 & 0.66 $\pm$ 0.01 & 4.1 $\pm$ 0.1 & 1.3 & 2.0 & 1.8$\times10^{5}$ & 2.4 \\ 
       & C$^{34}$S(1-0) & -40.3 $\pm$ 0.4 & 0.05 $\pm$ 0.01 & 3.2 $\pm$ 0.4 \\
       & SiO(J=1-0,v=0) & -41.1 $\pm$ 0.3 & 0.05 $\pm$ 0.01 & 3.7 $\pm$ 0.3 \\
       & HC$_{3}$N(5-4, F=4-3) & -39.6 $\pm$ 0.2 & 0.11 $\pm$ 0.01 & 3.8 $\pm$ 0.2 \\
\hline 
Bridge Core 2 & CS(1-0) & -32.5 $\pm$ 0.1 & 0.58 $\pm$ 0.01 & 3.3 $\pm$ 0.1 & 3.1 & 2.6 & 1.1$\times10^{5}$ & 4.5 \\ 
       & C$^{34}$S(1-0) & -32.0 $\pm$ 0.3 & 0.08 $\pm$ 0.01 & 2.5 $\pm$ 0.3 \\
       & SiO(J=1-0,v=0) & -30.7 $\pm$ 0.4 & 0.04 $\pm$ 0.01 & 3.2 $\pm$ 0.5 \\
       & HC$_{3}$N(5-4, F=4-3) & -31.3 $\pm$ 0.1 & 0.18 $\pm$ 0.01 & 2.0 $\pm$ 0.1 \\
\hline 
\end{tabular}

\begin{flushleft}
$^{a}$ An assumed distance, $d_{0}$ = 11 kpc was used for mass and density calculations. However, these values are easily scaled for an arbitrary distance, $d$, by multiplying by $(d/d_{0})^{2}$ and $(d/d_{0})^{-1}$ for mass and density respectively. \\
$^{b}$ The error in the calculated physical parameters are dominated by the statistical uncertainties associated with the abundance ratio of CS to molecular hydrogen. This uncertainty can be of the factor 2 (eg. \citealt{1987ASSL..134..561I}).
\end{flushleft}
\label{table:7mm_detections}
\end{table*}

\subsection*{7 mm emission in Group C}
Group C is a region towards the Galatic-eastern side of the gas structure located to the Galactic-west of HESS\,J1640$-$465 and HESS\,J1641$-$463 as traced in CO. Note that the extension of our 7 mm observations only reaches to include the region labelled ``1'' in the top panel of Figure \ref{fig:co_g338_g337}. We see large-scale extended and broad emission in the region in CS(1-0). This emission is in the same kinematic velocity ranges as that of the CO(1$-$0) emission. In Figure \ref{fig:co_g338_g337}, the average spectra shown for Extended region 1 include the CS(1$-$0) emission spectrum in red. Emission is seen between $\sim$ $-80$ to $-30$ km/s, similar to the profile seen in the $^{12}$CO and $^{13}$CO, and is likely tracing denser regions that exist inside cloud. There is also one dense core traced in CS which appears at $\sim -120$ km/s (dark blue in Figure \ref{fig:cs(1-0)_mom-3}).

\subsection*{7 mm emission in Group D}
Emission in CS(1-0) is seen in the Group D region slightly overlapping the Galactic-east side of the HESS\,J1641$-$463 contours in the \mbox{$\sim -25$ to $-20$ km/s} velocity range. It has a marginally extended morphology. The most intense emission, located towards the middle, is positionally coincident with a dense core detected in the NH$_{3}$(1-1) transition line by the H$_{2}$O southern Galactic Plane Survey (HOPS) \citep{2012MNRAS.426.1972P}.

A core of gas traced by CS emission is seen at $-80$ km/s (green in Figure \ref{fig:cs(1-0)_mom-3}) within the Galactic-north-east bounds of HESS\,J1641$-$463. This $-80$ km/s core appears marginally extended and is positionally coincident with detections made in NH$_{3}$(1,1) in the HOPS survey. Detections in the isotopologue transition C$^{34}$S(1-0) were also made at this position, as well as in SiO(J=1-0,v=0) and HC$_{3}$N(5-4, F=4-3). The average spectra of the 7 mm lines detected in an aperture centred at this core is displayed in the right panel of Figure \ref{fig:core_cs_spectra}, and we note that the emission in each line peaks at $-80$ km/s. A CH$_{3}$OH\,(I) maser is also seen positionally coincident with this core at $-79$ km/s. The velocity at which this core is detected suggests that it is not associated with the HII complex (which has v$_{\text{LSR}} \sim -30$ to $-40$ km/s).

\subsection{HI emission}
The atomic gas towards HESS\,J1640$-$465 and HESS\,J1641$-$463 was studied using HI data from the Southern Galactic Plane Survey (SGPS) \citep{2005ApJS..158..178M}. Integrated velocity maps were generated from the HI emission cubes over the same velocity intervals as the components traced in CO. Images of the HI integrated velocity maps can be found in the appendix Figure \ref{fig:sgps_big_moasic}. We note that a study of atomic gas using SGPS data towards HESS\,J1640$-$465 was carried out by \cite{2016A&A...589A..51S}, which focused on the velocity ranges from $-121$ to $-111$ km/s and from $-40$ to $-25$ km/s. We find that our results are similar to those presented towards HESS\,J1640$-$465.

There are prominent dips in the HI spectra towards the TeV sources which occur at velocities where emission features are seen in the CO spectra. These dips may be the result of HI self-absorption, caused by residual HI embedded in the CO. Figure\,\ref{fig:sgps_1641_dip} in the appendix is an example of such a dip in the spectrum towards HESS\,J1641$-$463 in component 1, and the corresponding intense emission that is seen in CO. To reduce the effect that these self-absorption dips may have on the calculation of the atomic gas parameters, and following the analysis technique in \cite{2012ApJ...746...82F}, we estimate the actual HI emission level by a linear interpolation connecting the adjacent shoulders of a dip. The dotted line in Figure \ref{fig:sgps_1641_dip} demonstrates this interpolation. This is a conservative estimate as the true spectrum is likely peaked, rather than just a straight line.

From the corrected spectra and, following \S3, mass and density estimates of the atomic gas contained in each component were calculated. Similar to the work by \cite{2016A&A...589A..51S}, we find that, when comparing the molecular gas traced by CO and the atomic gas traced by HI, the physical parameters of the atomic gas are a small fraction of molecular gas. The calculated masses of atomic gas in each component can be found in Table \ref{table:Kcr} in \S\ref{sec:discussion} and in Table\,\ref{table:all_kcr} in the appendix.

\section{Discussion}
\label{sec:discussion}
As mentioned previously in \S1, the production of $\gamma$-rays from the TeV sources HESS\,J1640$-$465 and HESS\,J1641$-$463 may be the result of hadronic scenarios, in which accelerated CRs are interacting with ambient gas, or leptonic scenarios, in which energetic electrons up-scatter background photons. We now consider the implications that our study of the interstellar gas towards these TeV sources have on the aforementioned origin scenarios.

\subsection{Hadronic scenarios}
\label{sec:hadronic_sce}

The segment of emission that includes components 1, 2, and 3 in the CO spectra traces gas that is at a velocity within $\sim10-20$ km/s of the reported systematic velocity of the HII regions in the HII complex (see \S\ref{sec:distance}). Assuming SNRs G338.3$-$0.0 and G338.5+0.1 and the TeV sources are all linked with this HII complex, the gas traced here is a potential target for accelerated CRs. The dense gas traced by CS at these velocities is found in a bridging region between the two TeV sources and may also be acting as CR target material.

A relationship to calculate the flux of $\gamma$-rays above a given energy level produced by hadronic interactions between CRs and CR-target material from the mass of the target material was derived by \cite{1991_Aharonian}. The expected $\gamma$-ray flux above some energy $E_{\gamma}$, assuming an $E^{-1.6}$ integral power law spectrum, is given by:

\begin{equation}
\label{eqn:gamma_from_gmc}
F(\geq E_{\gamma}) = 2.85\times10^{-13} E^{-1.6}_{\gamma} \left(\dfrac{M_{5}}{d^{2}_{kpc}}\right)k_{CR} \qquad \mathrm{cm^{-2}s^{-1}}
\end{equation}
where $M_{5}$ is the mass of the CR-target material in units of $10^{5}$ M$_{\odot}$, $d_{kpc}$ is the distance in kpc, $k_{CR}$ is the CR enhancement factor above that observed at Earth, and $E_{\gamma}$ is the minimum energy of $\gamma$-rays in TeV. 

From the results in \cite{2014MNRAS.439.2828A}, the $\gamma$-ray flux photon above 1 TeV towards HESS\,J1640$-$465 is determined to be $F(>1\text{TeV}) \sim 1.9 \times 10^{-12}$ cm$^{-2}$s$^{-1}$. From \cite{2014ApJ...794L...1A} the flux above 1 TeV towards HESS\,J1641$-$463 is $\sim 3.6 \times 10^{-13}$ cm$^{-2}$s$^{-1}$.

The total amount of CR-target material towards HESS\,J1640$-$465 and HESS\,J1641$-$463 is taken to be the sum of the molecular and atomic mass traced by emission in CO(1-0) and HI respectively. These masses, together with the calculated CR enhancement factors $k_{CR}$ for the total mass in components 1, 2, and 3 are displayed in Table \ref{table:Kcr}.  A complete list of $k_{CR}$ values for every gas component along the line of sight can be found in Table \ref{table:all_kcr} in the appendix. It should be noted that the $k_{CR}$ values here refer to \mbox{E $>$ 1 TeV} $\gamma$-rays and pertain only to higher energy CRs ($\gtrsim$ 10 TeV). Thus any CR energetics should be considered lower limits on the total CR energy. In addition, Equation \ref{eqn:gamma_from_gmc} assumes an $E^{-1.6}$ integral power law CR spectrum. This is different from the $\sim E^{-1.1}$ CR integral spectra needed to fit the $\gamma$-ray spectra of HESS\,J1640$-$465 and HESS\,J1641$-$463 \citep{2016A&A...589A..51S,2014ApJ...794L...1A}. By scaling Equation \ref{eqn:gamma_from_gmc} appropriately, the calculated CR enhancement factors would reduce by $\sim40\%$ for HESS\,J1640$-$465 and HESS\,J1641$-$463.

A similar study was conducted by \cite{2016A&A...589A..51S} on the physical properties of the ISM towards HESS\,J1640$-$465 using the same SGPS HI data, but with archival CO data from \cite{2001ApJ...547..792D}. The authors used different sized regions of integration and velocity ranges compared to those used here. Analyses utilising the same region size and velocity ranges on Mopra CO survey data return mass and densities parameters consistent with those presented in \cite{2016A&A...589A..51S}.

\begin{table*}
  \centering
   \caption{Calculated cosmic-ray enhancement valutes, k$_{\text{CR}}$, for the intrinsic Gaussian size of HESS\,J1640$-$465 and the maximum Gaussian extent of HESS\,J1641$-$463 for the gas related to components 1, 2, and 3 as defined in Figure \ref{fig:12co(1-0)_int}. Molecular  mass comes from CO analysis and atomic mass from HI analysis.}
\begin{tabular}{|c|c|c|c|c|c|c|c|}
\hline 
Region & Velocity range & Assumed distance & Molecular mass & Atomic mass & Total mass & k$_{\text{CR}}$ $^{a}$ \\ 
• & (km/s) & (kpc) & (M$_{\odot}$) & (M$_{\odot}$) & (M$_{\odot}$) & • & • \\ 
\hline 
HESS\,J1640-465 & -35 to -23 (Component 1) & 11.9 & 68,000 & 12,000 & 80,000 & 1,000 \\ 
              • & -45 to -35 (Component 2) & 11.2 & 47,000 & 9,300 & 56,000 & 1,400 \\ 
              • & -53 to -45 (Component 3) & 10.8 & 88,000 & 7,100 & 95,000 & 850 \\ 
    • & -53 to -23 (Components 1, 2, $\&$ 3) & 11.0 & 203,000 & 28,000 & 230,000 & 350 \\ 
\hline 
HESS\,J1641-463 & -35 to -23 (Component 1) & 11.9 & 97,000 & 5,000 & 102,000 & 150 \\ 
              • & -45 to -35 (Component 2) & 11.2 & 31,000 & 3,200 & 34,000 & 450 \\ 
              • & -53 to -45 (Component 3) & 10.8 & 11,000 & 2,100 & 13,000 & 1,200 \\ 
    • & -53 to -23 (Components 1, 2, $\&$ 3) & 11.0 & 139,000 & 10,000 & 149,000 & 100 \\ 
\hline 
\end{tabular} 
  \centering
\label{table:Kcr}
\begin{flushleft}
$^{a}$ Note that the CR enhancement factor, k$_{\text{CR}}$, is effectively independent of assumed distance as the distance terms in Eq. \ref{eqn:gamma_from_gmc} cancel with the distance assumptions for the mass calculations.
\end{flushleft}
\end{table*}

The CR enhancement factor k$_{\text{CR}}$ above 1 TeV towards HESS\,J1640$-$465 for the gas traced in components 1, 2, and 3 are of the order of $\sim 10^{3}$.  This value is consistent with a nearby (within a few pc) and young SNR ($\lesssim$ 5 kyr) such as G338.3$-$0.0 accelerating and injecting CRs into the ambient gas \citep{1996A&A...309..917A}. Thus, assuming the gas at either components 1, 2, or 3 are associated with the location of HESS\,J1640$-$465, a hadronic scenario is plausible. In a case where all of the gas traced in these components are summed and considered as CR target material associated with HESS\,J1640$-$465, the required $k_{CR}$ value becomes 350.

In the case of HESS\,J1641$-$463, the required $k_{CR}$ value for the molecular cloud positionally coincident (component\,1) is 150. Component 1 is dominant whereby summing the gas traced in components 2 and 3 marginally decreases the required $k_{CR}$ value. If the molecular cloud in component\,1 is indeed associated with HESS\,J1641$-$463, the hadronic scenario would be possible given its proximity with potential CR accelerators.

If the hadronic scenario holds true in both HESS\,J1640$-$465 and HESS\,J1641$-$463, then the total CR energy budget, $W_{p}$, can be given as $W_{p} = L_{\gamma} \tau_{pp}$, where $L_{\gamma}$ is the luminosity in $\gamma$-rays. $\tau_{pp}$ is the cooling time of protons through proton-proton collisions and is given by \citep{1996A&A...309..917A}: $\tau_{pp} \approx 6 \times 10^{7} (n/1 \text{cm}^{-3})^{-1}$ yr, where $n$ is the number density of the ambient gas.

HESS\,J1640$-$465 has a $\gamma$-ray luminosity of $L_{\gamma} = 9 \times 10^{34}$ erg s$^{-1}$ above 1 TeV at 11 kpc \citep{2014MNRAS.439.2828A}, while HESS\,J1641$-$463 has a luminosity of $4 \times 10^{34}$ erg s$^{-1}$ above 0.64 TeV at 11 kpc \citep{2014ApJ...794L...1A}. Thus $W_{p} \sim 10^{50}(n/1\text{cm}^{-3})^{-1}$ erg for HESS\,J1640$-$465, and $W_{p} \sim 10^{49}(n/1\text{cm}^{-3})^{-1}$ erg for HESS\,J1641$-$463. The number densities for both TeV sources presented in Table \ref{table:co_mass_param} in components 1, 2, and 3 are of the order $\sim 10^{2}$ cm$^{-3}$. $W_{p}$ is then $\sim10^{48}$ and $\sim10^{47}$ for HESS\,J1640$-$465 and HESS\,J1641$-$463 respectively, which is a fraction of the canonical amount of energy channelled into accelerated CRs by a SNR ($\sim10^{50}$ erg).

We note here that the hadronic modelling of HESS\,J1640$-$465 done in \cite{2016A&A...589A..51S} used the parametrisation of the $\gamma$-ray differential cross-section in the proton-proton interactions from \cite{2014PhRvD..90l3014K}. Using ambient proton densities of $\sim 10^{2}$ cm$^{-3}$, they found the total energy in accelerated protons to be \mbox{$\sim 10^{49} - 10^{50}$ erg}, consistent to first-order with a SNR scenario.

In one of the scenarios discussed in \cite{2014ApJ...794L...1A} and \cite{2015ApJ...812...32T}, CRs accelerated by the SNR G338.3$-$0.0, coincident with HESS\,J1640$-$465, have diffusively reached a molecular cloud coincident with HESS\,J1641$-$463. The centre of G338.3$-$0.0 lies $\sim 0.3^{\circ}$ from the far side of the maximum extent of HESS\,J1641$-$463, equivalent to $\sim 60$ pc at the assumed distance of 11 kpc. The maximum Gaussian extent of HESS\,J1641$-$463 is $0.05^{\circ}$ or $\sim$ 10 pc at this distance. Thus the filling factor of HESS\,J1641$-$463, assuming a spherical geometry with radius 10 pc, compared to the 60 pc radius sphere centred at SNR G338.3$-$0.0 is $\sim 0.005$. Assuming that $10^{50}$ erg was injected into accelerating CRs by G338.3$-$0.0, and assuming the CRs are uniformly distributed within the sphere, the total amount of energy in CRs at HESS\,J1641$-$463 is $\sim 5 \times 10^{47}$ erg. This value is consistent with the value of $W_{p}$ calculated above for HESS\,J1641$-$463, and thus the presented origin scenario is energetically plausible for the observed ISM.

The separation from the Galactic-eastern edge of the SNR G338.3$-$0.0 to the position of HESS\,J1641$-$463 is $\sim 0.15^{\circ}$ or 30 pc (at a distance of 11 kpc). The required CR enhancement factor of $k_{CR} \sim 100$ resulting from the diffusion of CRs from the SNR towards HESS\,J1641$-$463 is achievable according to \cite{1996A&A...309..917A}. Figure 1b of their paper shows the $k_{CR}$ values at several time epochs at a distance of 30 pc from a SNR. A $k_{CR}$ of $\sim 100$ is acquired at a source age between $10^{3}$ and $10^{4}$ years, similar to the estimated age of SNR G338.3$-$0.0. Their calculations assumed a source spectral index of 2.2 and that $D_{10} = 10^{26}$ cm$^{2}$s$^{-1}$, where $D_{10}$ is the diffusion coefficient when energy = 10 GeV. This value corresponds to slow diffusion and is not unexpected given the substantial amount of gas traced in the region between HESS\,J1640$-$465 and HESS\,J1641$-$463 in both CO and CS observations. This is because gas with larger values of $\overline{n}$ will have greater magnetic fields. The corresponding increase in the interaction between CR particles and magnetic fields would increase the rate of scattering, thereby decreasing the diffusion coefficient.

We can estimate the diffusion coefficient through the gas in this bridge region using Equation 2 from \cite{2007Ap&SS.309..365G}. The required value of the magnetic field inside the ISM is a function of $\overline{n}$ and is calculated following \cite{2010ApJ...725..466C}. In \S4 we have calculated the values of $\overline{n}$ for the diffuse and dense gas in the bridge region from CO and CS data respectively. These values have been presented in Tables \ref{table:co_mass_param} and \ref{table:7mm_detections}. 
For the diffuse CO traced gas, $B \sim 15 \mu\text{G}$ and \mbox{$D_{10} \sim \chi(4 \times 10^{27}$)} cm$^{2}$s$^{-1}$. For the dense CS traced cores, $B \sim 500 \mu\text{G}$ and $D_{10} \sim \chi(7 \times 10^{26}$) cm$^{2}$s$^{-1}$.
The parameter \mbox{$\chi < 1$} is a suppression factor that accounts for the suppression of the diffusion coefficient inside a turbulent cloud. For a moderate value of $\chi \sim 0.1$, the diffusion coefficient in this region would agree with a slow diffusion scenario. This diffusion scenario is a theoretically plausible explanation for the $\gamma$-ray emission from HESS\,J1641$-$463. The hardness of the TeV emission would be explained by higher energy protons preferentially reaching CR target material earlier and the effective exclusion of low energy CR due to the dense gas bridge.

As mentioned in \S4.1, emission in CO(1-0) traces an extended molecular cloud structure to the Galactic-west of HESS\,J1640$-$465. The angular separation of this cloud is comparable to the separation between SNR G338.3$-$0.0 and HESS\,J1641$-$463 ($\sim0.3^{\circ}$). If the scenario in which CRs escaping from SNR G338.3$-$0.0 are generating the TeV emission of HESS\,J1641$-$463 is true, then one might have expected this other giant molecular cloud to glow in $\gamma$-rays. However, while at similar angular separations from the SNR, this molecular cloud may be at different distance along the line-of-sight. The CO(1$-$0) and CS(1$-$0) emission (see Figure \ref{fig:co_g338_g337}) is very broad and is seen between $\sim-80$ and $-20$ km/s in velocity. This is in contrast with the emission from molecular cloud seen towards HESS\,J1641$-$463 which is seen at $\sim-30$ km/s. Differences between the Galactic-rotation curve solution at these velocities range up to several kpc. Since an increase in distance of even $\sim100$ pc from the source SNR would drastically diminish the available CRs \citep{1996A&A...309..917A}, it is possible that CRs may not have yet reached this molecular cloud. In addition, CRs escaping from an accelerator can diffuse anisotropically, tending to propagate along magnetic field lines (eg. \cite{2013MNRAS.429.1643N}). It is possible that the orientation of the magnetic fields in this region are directing CRs away from this molecular cloud.

An alternate scenario is that SNR\,G338.5+0.1, coincident with HESS\,J1641$-$463, is accelerating CRs that are interacting with the ISM. Modelling by \cite{2014ApJ...794L...1A} indicate that the hard proton spectrum required to generate the $\gamma$-ray emission agrees well with CRs accelerated by a young SNR. The critical factor is then the age of SNR G338.5+0.1. A 5-17 kyr middle aged SNR \citep{2014ApJ...794L...1A} would disfavour this scenario, and lend support to the diffusion scenario discussed above. In either case, the molecular cloud found in component 1 of our study towards HESS\,J1641$-$463 provides ample target material for accelerated CRs to interact with, producing $\gamma$-rays through the hadronic channel.

\subsection{Leptonic Scenarios}
We now consider the leptonic scenarios, in which TeV emission is primarily due to accelerated electrons interacting with ambient photons via the inverse-Compton effect, for HESS\,J1640$-$465 and HESS\,J1641$-$463 in light of our ISM study.

The leptonic scenario for HESS\,J1640$-$465 has been developed and explored by several previous works \citep{2007ApJ...662..517F,2009ApJ...706.1269L,004.2936v2,2014MNRAS.439.2828A,2014ApJ...788..155G}. In particular, the scenario involves the electrons being accelerated at the termination shock of a PWN near the centroid of the TeVs source which is powered by PSR\,J1640$-$4631 \citep{2014ApJ...788..155G}. In our ISM study and considerations in the previous section, we have shown that there is sufficient target material towards HESS\,J1640$-$465 for a purely hadronic origin, given a local CR accelerator such as SNR G338.3$-$0.0. As such, our study does not rule out either model, and it possible that the TeV emission from HESS\,J1640$-$465 has contributions from both leptonic and hadronic processes, although some fine-tuning is required to explain the smooth power-law spectrum seen by Fermi-LAT \citep{2014ApJ...794L..16L}.

A leptonic origin has been previously considered for HESS\,J1641$-$463 \citep{2014ApJ...794L...1A} but was strongly disfavoured, stemming from the lack of a characteristic break in the $\gamma$-ray spectrum and the extreme difficulty in accelerating a population of electrons to the required energies. For completeness, however, we now consider a leptonic diffusion scenario in which CR electrons, perhaps from the PSR J1640$-$4631, are diffusing through the gas bridge between HESS\,J1640$-$465 and HESS\,J1641$-$463. 

The cooling time of CR electrons due to synchrotron radiation can be given by $\tau_{\text{sync}} \approx (b_{s}\gamma)^{-1}$ s, where $b_{s} = 1.292 \times 10^{-15} (B/\text{mG})^{2}$ s$^{-1}$, and the diffusion time of CRs over a distance $d$ is given by $\tau_{\text{diff}} = d^{2}/(6D(E))$, where $D(E)$ is the diffusion coefficient at energy $E$ \citep{1964ocr..book.....G}. We consider the case where E$_{e}$ = 5 TeV. At this energy, $\gamma$-rays produced via inverse-Compton scattering would have energies $\sim 200$ GeV (in the Thompson regime) which is near the lower limit of detectability by H.E.S.S.

Dense cores of gas within the bridge region are traced in CS observations with $\overline{n} \sim 10^{5}$ cm$^{-3}$. These cores have a diameter of $\sim 2'$, which corresponds to $\sim 6$ pc at a line-of-sight distance of 11 kpc. The magnetic field in these cores are calculated following \cite{2010ApJ...725..466C}, using the values of $\overline{n}$ presented in \S4. Using this, $\tau_\text{{sync}}$ and $\tau_{\text{diff}}$ were found to be \mbox{$\sim 10$ yr} and \mbox{$\sim1$ kyr} respectively. This implies that CR electrons would rapidly lose their energy via synchrotron losses, and be unable to pass through the dense cores. For completeness, we note that calculations using emission from the diffuse gas tracers $^{12}$CO and $^{13}$CO in these core regions give $\overline{n} \sim 4\times10^{3}$ cm$^{-3}$.

In an optimistic scenario, CR electrons could have paths that avoid the dense cores while diffusing through the gas bridge region. From \S4, the diffuse gas traced by CO observations across the whole bridge region have \mbox{$\overline{n} \sim 5\times10^{2}$ cm$^{-3}$}. Assuming that the distance across the gas bridge is $\sim 30$ pc, $\tau_{\text{sync}}$ and $\tau_{\text{diff}}$ are calculated to be $\sim 11$ kyr and $\sim 4$ kyr respectively. Hence, in an optimistic scenario, some electrons would be able to diffuse through the gas bridge. However, realistically this is somewhat unlikely given the prevalence of dense gas in the region, as traced by CS emission. CR electrons would be effectively blocked by the dense regions of the gas bridge and be unable to generate $\gamma$-rays which may be contributing to HESS\,J1641$-$463.

\section{Conclusions}
In this paper we have used data collected by the Mopra Radio Telescope in the 3 mm and 7 mm wavelengths as well as archival HI data to investigate the molecular and atomic gas towards the VHE $\gamma$-ray sources HESS\,J1640$-$465 and HESS\,J1641$-$463. The gas investigated here may be target material for accelerated CRs, producing TeV $\gamma$-rays via hadronic interactions.

CO(1-0) observations from the Mopra Galactic Plane Survey revealed multiple diffuse molecular gas components at numerous velocities along the line-of-sight positionally coincident with both TeV sources. In particular, substantial detections were made at velocities within $\sim 10-20$ km/s of the reported systematic velocity of the HII region ($-32$ km/s). The gas traced in Components 1, 2, and 3 as described in \S4 ($-53$ to $-23$ km/s) may be then associated with the SNRs, HII region and the VHE $\gamma$-ray sources. Of particular note is the molecular cloud traced in Component 1 positionally coincident with HESS\,J1641$-$463.

7 mm observations in the CS(1-0) lines revealed a region of dense gas cores coincident with intense emission in the CO(1-0) lines. This gas formed a ``bridge" of material located between the two TeV $\gamma$-ray sources.

Mass and density estimates derived from CO, CS, and HI for gas components towards HESS\,J1640$-$465 and HESS\,J1641$-$463 allowed for an investigation of the available CR target mass. Assuming the total gas mass in Components 1, 2, or 3 towards HESS\,J1640$-$465 is CR target material in a hadronic scenario for TeV $\gamma$-ray production, the required $W_{p}$ is $\sim 10^{48}$ erg and the required CR density would be of the order of $\sim10^{3}$ times that seen at Earth. For HESS\,J1641$-$463 if the molecular cloud positionally coincident traced in Component 1 is CR target material, then the required $W_{p}$ is $\sim 10^{47}$ erg, and the required CR density would be of the order of $\sim10^{2}$ times that seen at Earth.

We also investigated the scenario in which TeV emission from HESS\,J1641$-$463 is due to high energy CRs from SNR G338.0-0.0, coincident with HESS\,J1640$-$465, diffusively reaching CR target material seen in our data. We find that the scenario is a plausible explanation which readily explains the hardness of the TeV emission from HESS\,J1641$-$463. We do not, however, discount the scenario in which SNR G338.5+0.1 coincident with HESS\,J1641$-$463, is providing the required CRs.

A scenario in which CR electrons from PSR J1640$-$4631 were diffusing towards HESS\,J1641$-$463 was considered. However, it is somewhat unlikely due to dense cores of gas present in the bridge between HESS\,J1640$-$465 and HESS\,J1641$-$463 which would effectively block the path of the electrons. This is in addition to arguments presented by \cite{2014ApJ...794L...1A} that disfavour a leptonic origin.

Future $\gamma$-ray measurements taken by next-generation ground based $\gamma$-ray telescopes systems (eg. Cherenkov Telescope Array), will have greatly increased sensitivity above 10 TeV and have angular resolutions similar to that in this study of the interstellar gas. This will allow more detailed morphological comparison between the TeV $\gamma$-ray emission and gas and allow a deeper investigation of the nature of HESS\,J1640$-$465 and HESS\,J1641$-$463.

\section*{Acknowledgements}
The Mopra radio telescope is part of the Australia Telescope National Facility. Operations support was provided by the University of New South Wales and the University of Adelaide. The UNSW Digital Filter
Bank used for the observations with Mopra was provided with financial support from the Australian Research Council (ARC), UNSW, Sydney and Monash universities. We also acknowledge ARC support through Discovery Project DP120101585.
Sabrina Casanova acknowledges the support of the Polish Science Centre through the Opus grant UMO-2014/13/B/ST9/00945.

\bibliography{bibliography_clean}
\appendix
\section{}

\begin{figure*}
  \centering
    \includegraphics[width=0.85\textwidth]{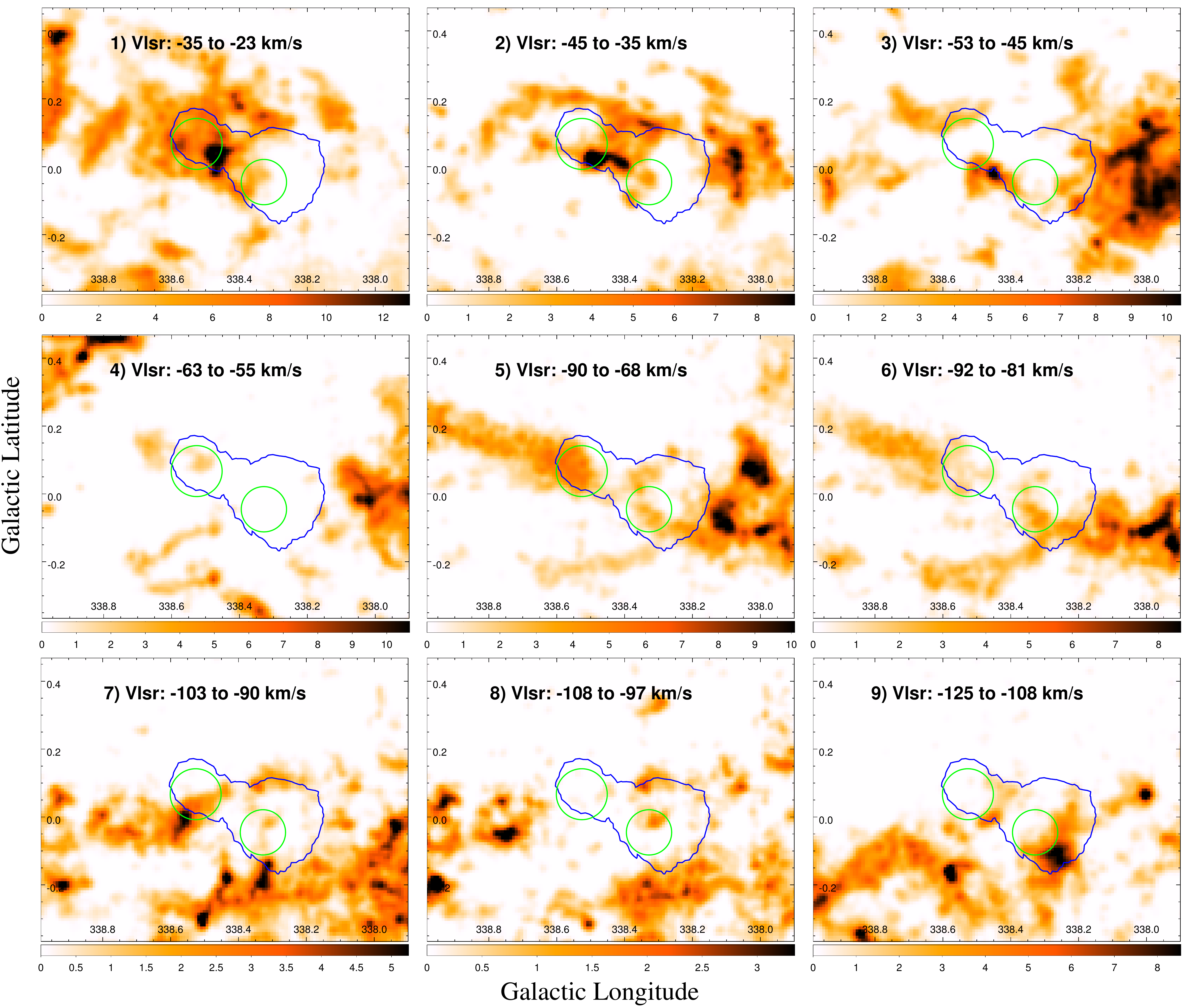}
    \rule{40em}{0.5pt}
  	\caption[what]{Integrated $^{13}$CO(1-0) emission images [K km/s] over indicated velocity intervals. Single blue $5\sigma$ significance H.E.S.S. contour used for clarity and illustration purposes. The position and extent of SNR G338.5+0.1 and SNR G338.3-0.0 are indicated by the left and right solid green circles respectively in each panel.}
  \label{fig:13co_big_moasic}
\end{figure*}

\begin{table*}
  \centering
  \caption{$^{13}$CO(1-0) line parameters, and the corresponding calculated gas parameters, from the apertures as indicated in Figure \ref{fig:12co(1-0)_int}. Calculations were made following \S3.1, but using the $^{13}$CO(1-0) X-factor, $X_{^{13}\text{CO(1-0)}} = 4.92 \times 10^{20}$ (K km/s)$^{-1}$ \citep{2001ApJ...551..747S}. Masses and density have been scaled to account for an additional 20$\%$ He component.}
  \begin{tabular}{|c|c|c|c|c|c|}
  \hline 
  Velocity range & Region  & Distance$^{a}$ & $\overline{N_{H_{2}}}$ $^{b}$ & Mass $^{b}$ & $\overline{n}$ $^{b}$ \\ 
  (km/s) &    & (kpc) & ($10^{21}$ cm$^{-2}$) & (M$_{\odot}\times10^{4}$) & ($10^{2}$ cm$^{-3}$) \\
  \hline 
  -35 to -23 & HESS\,J1640$-$465    & 11.9 & 1.7 & 2.7  & 0.8 \\

  (Component 1)& HESS\,J1641$-$463  & 11.9 & 6.7 & 5.1  & 4.4 \\
  
    & Bridge                        & 11.9 & 7.1 & 10.1 & 3.3 \\ 
  \hline
  -45 to -35 & HESS\,J1640$-$465    & 11.2 & 2.9 & 4.3  & 1.4 \\ 

  (Component 2) & HESS\,J1641$-$463 & 11.2 & 3.0  & 2.0  & 2.1 \\
  
    & Bridge                        & 11.2 & 8.7  & 11.7  & 4.3 \\ 
  \hline
  -53 to -45 & HESS\,J1640$-$465    & 10.8 & 1.9  & 2.5  & 0.9 \\ 

  (Component 3) & HESS\,J1641$-$463 & 10.8 & 0.6  & 0.4  & 0.4 \\ 

    & Bridge                        & 10.8 & 3.9 & 4.9 & 2.0 \\ 

  \hline
\end{tabular}
\begin{flushleft}
$^{a}$ Assumed distances, $d_{0}$, used for mass and density calculations are derived from the Galactic rotation curve presented in \cite{2007A&A...468..993K}. However, these values are easily scaled for an arbitrary distance, $d$, by multiplying by $(d/d_{0})^{2}$ and $(d/d_{0})^{-1}$ for mass and density respectively.\\
$^{b}$ The error in the calculated physical parameters are dominated by the statistical uncertainties associated with the $^{13}$CO to H$_{2}$ conversion factor ($X_{^{13}\text{CO(1-0)}}$) and is on the order of 30$\%$.
\end{flushleft}
\label{table:13co_mass_param}
\end{table*}

\begin{figure*}
  \centering
    \includegraphics[width=0.85\textwidth]{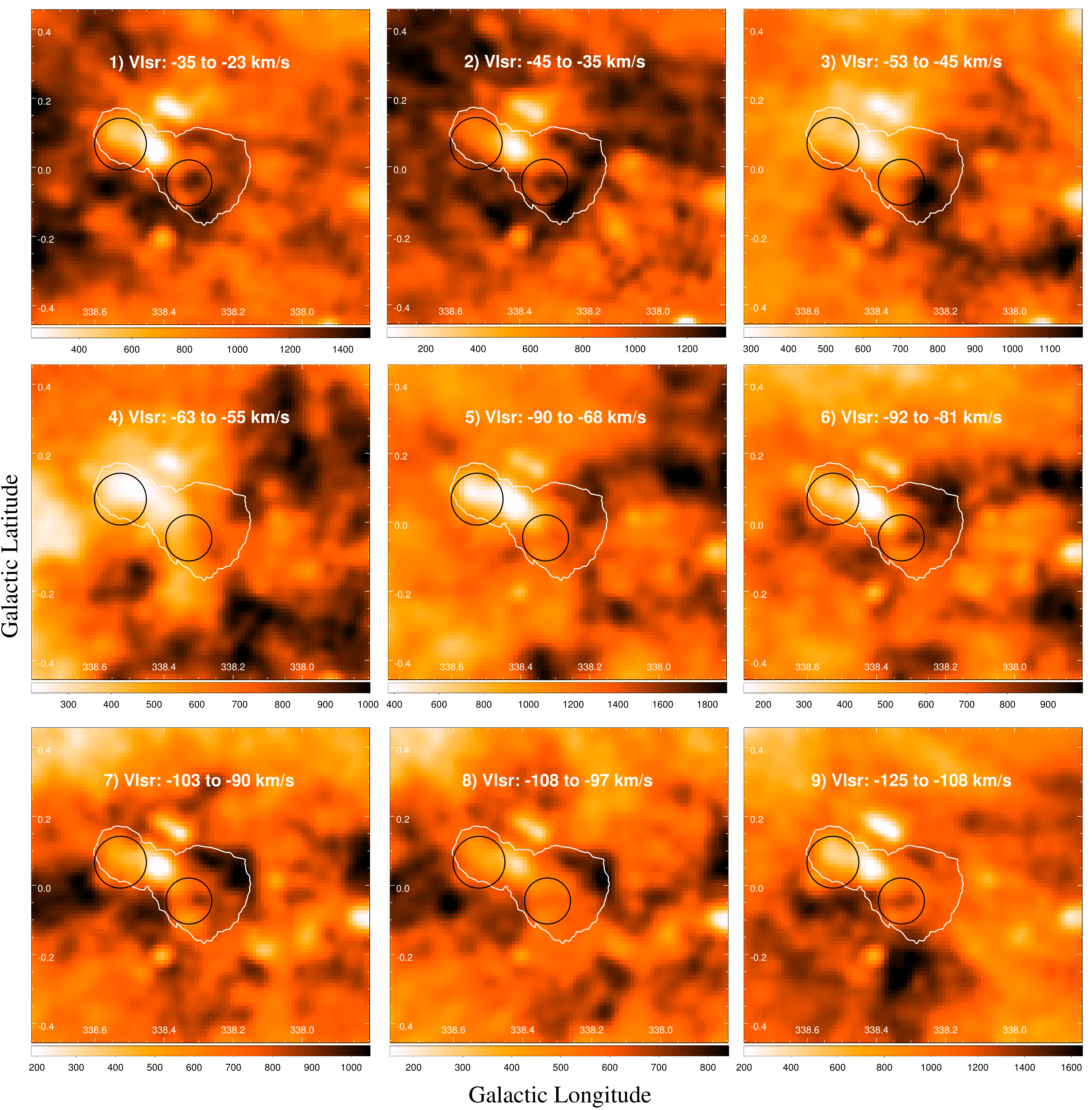}
    \rule{40em}{0.5pt}
  	\caption[what]{Integrated HI emission images [K km/s] from SGPS data over indicated velocity intervals. Single white $5\sigma$ significance H.E.S.S. contour used for clarity and illustration purposes. The position and extent of SNR G338.5+0.1 and SNR G338.3-0.0 are indicated by the left and right solid black circles respectively in each panel.}
  \label{fig:sgps_big_moasic}
\end{figure*}

\begin{figure*}
  \centering
  \vspace{0.5cm}
    \includegraphics[width=0.5\textwidth]{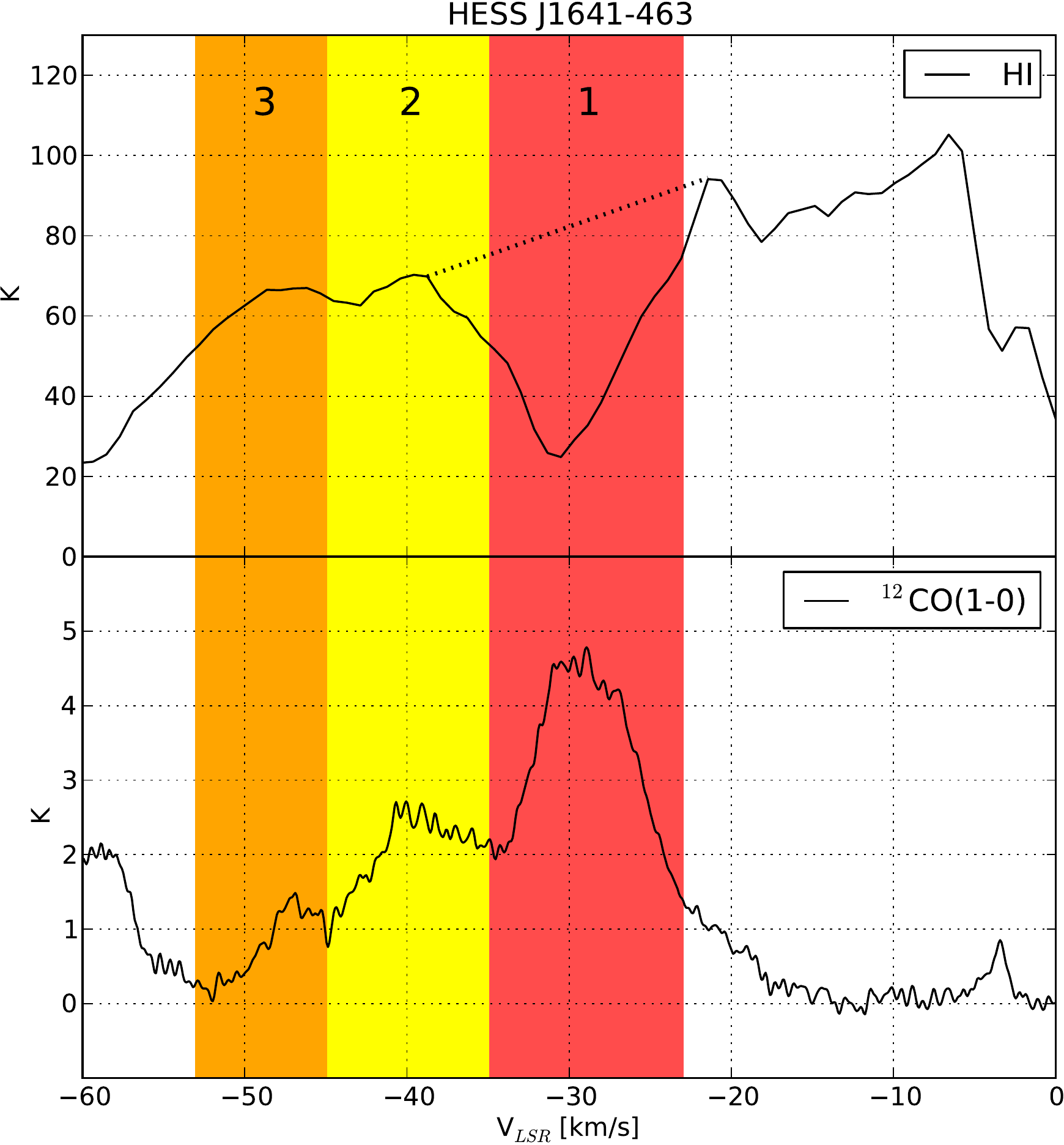}
    \rule{40em}{0.5pt}
  	\caption[what]{Average HI and $^{12}$CO(1-0) emission spectrum towards HESS\,J1641$-$463 from the SGPS and Mopra CO survey respectively. A possible example of a self-absorption dip in the HI is seen at $\sim$-30 km/s in component 1, where a corresponding peak exists in the CO. The dashed line is the linear interpolation used to estimate the true HI emission.}
  \label{fig:sgps_1641_dip}
\end{figure*}

\begin{table*}
  \centering
   \caption{Calculated cosmic-ray enhancement values, k$_{\text{CR}}$, for the intrinsic Gaussian size of HESS\,J1640$-$465 and the maximum Gaussian extent of HESS\,J1641$-$463 for the gas related to all individual components as defined in Figure \ref{fig:12co(1-0)_int}. Molecular  mass comes from CO analysis and atomic mass from HI analysis.}
\begin{tabular}{|c|c|c|c|c|c|c|c|}
\hline 
Region & $v_{\text{LSR}}$ range & Assumed distance $^{a}$ & Molecular mass & Atomic mass & Total mass & k$_{\text{CR}}$ $^{b}$ \\ 
• & (km/s) & (kpc) & (M$_{\odot}$) & (M$_{\odot}$) & (M$_{\odot}$) & • & • \\ 
\hline 
HESS\,J1640-465 & -35 to -23 (Component 1) & 11.9 & 68,000 & 12,000 & 80,000 & 1,000 \\ 
• & -45 to -35 (Component 2) & 11.2 & 47,000 & 9,300 & 56,000 & 1,400 \\ 
• & -53 to -45 (Component 3) & 10.8 & 88,000 & 7,100 & 95,000 & 850 \\ 
• & -63 to -55 (Component 4) & 10.4 & 15,000 & 4,400 & 19,000 & 3,800  \\ 
• & -90 to -68 (Component 5) & 9.6 & 58,000 & 8,000 & 64,000 & 950\\ 
• & -92 to -81 (Component 6) & 9.3 & 39,000 & 4,200 & 43,000 & 1300\\ 
• & -103 to -90 (Component 7) & 8.9 & 20,000 & 4,300 & 24,000 & 2210\\ 
• & -108 to -97 (Component 8) & 8.7 & 15,000 & 3,000 & 18,000 & 2800\\ 
• & -125 to -108 (Component 9) & 8.15 & 40,000 & 4,700 & 45,000 & 990\\ 
\hline 
HESS\,J1641-463 & -35 to -23 (Component 1) & 11.9 & 97,000 & 5,000 & 102,000 & 150 \\ 
• & -45 to -35 (Component 2) & 11.2 & 31,000 & 3,200 & 34,000 & 450 \\ 
• & -53 to -45 (Component 3) & 10.8 & 11,000 & 2,100 & 13,000 & 1,200 \\ 
• & -63 to -55 (Component 4) & 10.4 & 17,000 & 1,000 & 18,000 & 750\\ 
• & -90 to -68 (Component 5) & 9.6 & 37,000 & 1,700 & 39,000 & 300\\ 
• & -92 to -81 (Component 6) & 9.3 & 15,000 & 1,100 & 16,000 & 680\\ 
• & -103 to -90 (Component 7) & 8.9 & 13,000 & 1,400 & 14,000 & 720\\ 
• & -108 to -97 (Component 8) & 8.7 & 5,500 & 1,200 & 7,000 & 1,400\\ 
• & -125 to -108 (Component 9) & 8.15 & 7,100 & 1,300 & 8,000 & 1,000 \\ 
\hline
\label{table:all_kcr}
\end{tabular} 
\begin{flushleft}
$^{a}$The assumed distance was calculated using the average $v_{\text{LSR}}$ in the interval with the Galactic rotation curve presented by \cite{2007A&A...468..993K}. For consistency we have used the far distance solutions for every component. Masses can be scaled for an arbitrary distance, $d$, by multiplying by $(d/d_{0})^{2}$ where $d_{0}$ is the assumed distance.

$^{b}$ The CR enhancement factor, k$_{\text{CR}}$, was calculated following Equation 10 from \cite{1991_Aharonian} considering the total masses presented in this table as CR-target material. Note also that k$_{\text{CR}}$ is independent of assumed distance as the distance terms the equation cancel with the distance assumptions for the mass calculations.
\end{flushleft}

\label{table:Kcr_appendix}
\end{table*}

\label{lastpage}
\end{document}